\documentclass[aip,jcp,preprint,groupedaddress]{revtex4-1}%
\usepackage{graphicx}
\usepackage{tabularx}
\usepackage{dcolumn}
\usepackage{longtable}
\usepackage{tensor}
\usepackage{color}
\usepackage{placeins}
\usepackage{bm}
\usepackage{amsmath}
\usepackage{amsfonts}
\usepackage{amssymb}
\usepackage{float}%
\providecommand{\U}[1]{\protect\rule{.1in}{.1in}}
\begin{document}
\title{On-the-fly \emph{ab initio} semiclassical evaluation of time-resolved
electronic spectra}
\author{Tomislav Begu\v{s}i\'{c}}
\author{Julien Roulet}
\author{Ji\v{r}\'i Van\'i\v{c}ek}
\email{jiri.vanicek@epfl.ch.}
\affiliation{Laboratory of Theoretical Physical Chemistry, Institut des Sciences et
Ing\'enierie Chimiques, Ecole Polytechnique F\'ed\'erale de Lausanne (EPFL),
CH-1015, Lausanne, Switzerland}
\date{\today}

\begin{abstract}
We present a methodology for computing vibrationally and time-resolved
pump-probe spectra, which takes into account all vibrational degrees of
freedom and is based on the combination of the thawed Gaussian approximation
with on-the-fly \textit{ab initio} evaluation of the electronic structure. The
method is applied to the phenyl radical and compared with two more approximate
approaches based on the global harmonic approximation---the global harmonic
method expands both the ground- and excited-state potential energy surfaces to
the second order about the corresponding minima, while the combined global
harmonic/on-the-fly method retains the on-the-fly scheme for the excited-state
wavepacket propagation. We also compare the spectra by considering their means
and widths, and show analytically how these measures are related to the
properties of the semiclassical wavepacket. We find that the combined approach
is better than the global harmonic one in describing the vibrational
structure, while the global harmonic approximation estimates better the
overall means and widths of the spectra due to a partial cancellation of
errors.Although the full-dimensional on-the-fly \emph{ab initio} result seems
to reflect the dynamics of only one mode, we show, by performing exact quantum
calculations, that this simple structure cannot be recovered using a
one-dimensional model. Yet, the agreement between the quantum and
semiclassical spectra in this simple, but anharmonic model lends additional
support for the full-dimensional \emph{ab initio} thawed Gaussian calculation
of the phenyl radical spectra. We conclude that the thawed Gaussian
approximation provides a viable alternative to the expensive or unfeasible
exact quantum calculations in cases, where low-dimensional models are not
sufficiently accurate to represent the full system.

\end{abstract}
\maketitle

\graphicspath{{"d:/Group Vanicek/Desktop/PP_TGA/figures/"}
{./figures/}{C:/Users/Jiri/Dropbox/Papers/Chemistry_papers/2018/PP_TGA/figures/}}

\section{\label{sec:intro}Introduction}

Vibrationally resolved electronic spectroscopy provides a powerful tool for
studying both electronic and nuclear motions in molecular systems. Among other
things, this experimental technique gives insight into the photoinduced
molecular dynamics and helps elucidate the shape of molecular potential energy
surfaces. Its steady-state version, used for many years, has been interpreted
mostly through the time-independent Franck--Condon picture, but an
alternative, time-dependent
approach,\cite{Heller:1981a,Mukamel:1995,Tannor:2007} which examines the
evolution of the nuclear wavepacket before evaluating the spectra, is equally
valid and, indeed, more fundamental. With the advent of ultrafast
spectroscopy, this time-dependent approach became much more popular because it
is a natural way to interpret time-resolved spectra. The best-known
experimental setup for measuring time-resolved spectra is the pump-probe
technique, consisting of a pump laser pulse, which initiates a dynamical
process in the system and is followed by a probe pulse, whose response is
measured. Advances in the field, namely the development of broadband setups,
which use ultrashort optical pulses, led to the direct observation of nuclear
wavepacket dynamics in various
systems.\cite{Fragnito_Shank:1989,Polli_Cerullo:2007,Polli_DeSilvestri:2008,Veen_Chergui:2011,Monni_Chergui:2017}
These spectra are commonly analyzed with fitting procedures to recover the
kinetics,\cite{Ernsting_Farztdinov:2001,VanStokkum_VanGrondelle:2004} with
Fourier transformation to identify the main frequency
components,\cite{Liebel_Kukura:2013} or with simple low-dimensional models to
simulate the vibronic features.\cite{Veen_Chergui:2011} However,
computationally feasible full-dimensional methods for simulating vibrationally
and time-resolved spectra of large polyatomic molecules are missing.

The evaluation of vibrationally resolved electronic spectra is often limited
to approximate potential energy surfaces, e.g., global harmonic
models,\cite{Baiardi_Barone:2013} or exact quantum calculations taking into
account only a few degrees of freedom. To include anharmonicity effects
without having to evaluate the global potential energy surfaces, a number of
trajectory-based on-the-fly \emph{ab initio} (also called \textquotedblleft
direct dynamics\textquotedblright\ or \textquotedblleft first-principles
dynamics\textquotedblright) methods have been developed, ranging from Gaussian
basis methods, such as \emph{ab initio} multiple
spawning,\cite{Curchod_Martinez:2018} coupled coherent
states,\cite{Shalashilin_Mark:2004} variational multiconfigurational
Gaussians,\cite{Richings_Lasorne:2015}, multiconfigurational
Ehrenfest,\cite{Saita_Shalashilin:2012} and Gaussian dephasing
representation,\cite{Sulc_Vanicek:2013} to semiclassical methods based on the
Herman-Kluk propagator\cite{Herman_Kluk:1984,Miller:2001}, including versions
approximating the
prefactor\cite{Tatchen_Pollak:2009,Ianconescu_Pollak:2013,Wong_Roy:2011} or
the time-averaged semiclassical initial-value
representation,\cite{Ceotto_Atahan:2009b} and its multiple coherent
states\cite{Gabas_Ceotto:2017} or divide-and-conquer
extensions.\cite{Ceotto_Conte:2017} Despite their success, these methods are
rather expensive when it comes to computing the vibrational structure of
electronic spectra, since they require a large number of classical
trajectories or basis functions to converge. As an alternative, we have
recently implemented an on-the-fly \emph{ab initio}
version\cite{Wehrle_Vanicek:2014,Wehrle_Vanicek:2015} of the single-trajectory
thawed Gaussian approximation\cite{Heller:1975} as a computationally efficient
compromise between rather expensive multiple-trajectory semiclassical methods
and cheap, but restrictive and often flawed, global harmonic methods. The
thawed Gaussian approximation, based on propagating a single Gaussian
wavepacket, is commonly regarded as a starting point for developing more
accurate approaches, such as the exact wavepacket propagation of Lee and
Heller\cite{Lee_Heller:1982}, generalized Gaussian wavepacket
dynamics,\cite{Huber_Heller:1987} off-center-guided
wavepackets,\cite{Heller:2006} Hagedorn wavepacket
propagation,\cite{Faou_Lubich:2009} time-sliced thawed Gaussian
propagation,\cite{Kong_Batista:2016} and coupled wavepackets for nonadiabatic
molecular dynamics.\cite{White_Mozyrsky:2016} In the hybrid semiclassical
approach of Grossmann,\cite{Grossmann:2006,Buchholz_Ceotto:2018} the original
thawed Gaussian approximation is applied to model the \textquotedblleft
harmonic\textquotedblright\ degrees of freedom, while treating the rest by
more sophisticated semiclassical methods. In this context, the success of the
on-the-fly \emph{ab initio} thawed Gaussian approximation in reproducing
steady-state absorption, emission, and photoelectron spectra of rather large
systems\cite{Wehrle_Vanicek:2014,Wehrle_Vanicek:2015} is somewhat surprising.
This motivated us to further extend \textit{ab initio} thawed Gaussian
approximation to include the non-Condon effects, e.g., in symmetry-forbidden
electronic transitions, where the initial wavepacket is not a Gaussian, but a
Gaussian multiplied by a polynomial.\cite{Lee_Heller:1982} In all cases
studied so far, the methods based on the on-the-fly \emph{ab initio} thawed
Gaussian approximation proved to be more accurate than those based on the
global harmonic models.

Here, we adapt the on-the-fly \emph{ab initio} thawed Gaussian approximation
in order to evaluate vibrationally and time-resolved electronic spectra, and
compare it with the commonly used adiabatic global harmonic approach, which
approximates the two potential energy surfaces to the second order about the
corresponding minima. Furthermore, we propose and validate a \textquotedblleft
combined\textquotedblright\ global harmonic/on-the-fly method, which
approximates the ground potential energy surface with a harmonic potential,
but propagates the nuclear wavepacket on the excited state with the on-the-fly
\emph{ab initio} thawed Gaussian approximation. The properties of the
wavepackets propagated using the on-the-fly, global harmonic, and combined
approaches are analyzed through the differences in the spectra they generate.
In particular, we discuss the effect of anharmonicity in both electronic states.

As a proof of principle, we apply the methods to the phenyl radical, a species
studied for its rich photochemistry induced by visible and ultraviolet light.
Higher electronic states of phenyl radical undergo photodissociation through
multiple pathways, as shown by various
authors.\cite{Song_Brazier:2012,Mebel_Landera:2012,Cole-Filipiak_Neumark:2014}
Here, we focus our attention on the first excited state $\tilde{\text{A}}%
^{2}\text{B}_{1}$, which does not undergo photodissociation and is responsible
for the visible spectrum of the phenyl radical. The $n\leftarrow\pi$ character
of the $\tilde{\text{A}}^{2}$B$_{1}\leftarrow\tilde{\text{X}}^{2}$A$_{1}$
transition from the ground to the first excited state is in agreement with
both experimental and theoretical
studies.\cite{Biczysko_Barone:2009,Freel_Heaven:2011} The high-resolution
cavity ring-down spectroscopic study of the phenyl radical predicted a planar
geometry of the excited electronic state, involving an expansion of the carbon
ring, and excited-state lifetime of 96 ps.\cite{Freel_Heaven:2011} The
vibronic band corresponding to this transition, first characterized by Porter
and Ward\cite{Porter_Ward:1965} and later by
Radziszewski,\cite{Radziszewski:1999} exhibits a rich structure arising from
the considerable differences between the potential energy surfaces of the
ground and excited electronic states. \emph{Ab initio}
calculations\cite{Biczysko_Barone:2009,Baiardi_Barone:2013} found strong
coupling between the modes, also known as the Duschinsky rotation or mixing,
and a large displacement of multiple normal modes\cite{Patoz_Vanicek:2018,
Begusic_Vanicek:2018}---the two most displaced modes correspond to the
in-plane carbon ring vibrations (shown in Fig.~1 of the Supporting
Information). Due to their significant displacement, the anharmonicity of the
potential affects the dynamics and the resulting spectrum, as demonstrated
using the on-the-fly \emph{ab initio} semiclassical
dynamics.\cite{Patoz_Vanicek:2018, Begusic_Vanicek:2018} In addition, we have
recently analyzed the Herzberg--Teller contribution, due to the dependence of
the electronic transition dipole moment on nuclear coordinates, to the
steady-state $\tilde{\text{A}}^{2}\text{B}_{1}\leftarrow\tilde{\text{X}}%
^{2}\text{A}_{1}$ electronic spectrum of the phenyl
radical:\cite{Patoz_Vanicek:2018, Begusic_Vanicek:2018} by comparing the
spectra evaluated using the Condon and Herzberg--Teller approximations, we
have shown that the Herzberg-Teller correction to the transition dipole moment
does not affect the spectrum significantly, justifying the Condon
approximation, which will be also employed below for the time-resolved spectra.

Because the time-resolved electronic spectra of phenyl radical have not yet
been measured, our calculations provide a prediction of the experimental
signal corresponding to the time-resolved stimulated emission (and
ground-state bleach). Moreover, our calculations, although not yet confirmed
by an experiment, suggest new ways of analyzing time-resolved spectra. Spectra
with low vibrational resolution can be described well by their position and
width, which are measured by the mean and standard deviation of the normalized
spectral lineshape\cite{Pollard_Mathies:1990a,Ferrer_Santoro:2013}---we show
how the mean frequency and width of the spectra are directly related to the
properties of the thawed Gaussian wavepacket and can be evaluated using a
single classical trajectory on the excited electronic surface.

Clearly, the single-Gaussian ansatz imposes a severe constraint on the shape
of the nuclear wavepacket, making the thawed Gaussian approximation
susceptible to failure in the presence of nonadiabatic effects or wavepacket
splitting. In such situations, one is forced to use exact quantum methods or
semiclassical methods based on multiple guiding trajectories, such as those
mentioned earlier in this Section, or multiple thawed Gaussian wavepackets. In
addition, the on-the-fly \emph{ab initio} version of the thawed Gaussian
approximation has only been validated on linear spectra, which depend on the
wavepacket autocorrelation function. Such spectra, unless highly-resolved, can
validate the time-dependent wavepacket only at times when it returns to the
Franck--Condon region because linear spectra depend weakly on the form of the
wavepacket outside this region. In general, the approximation is expected to
be qualitatively correct in single-well potentials, where the wavepacket does
not split during the short-time propagation needed to describe electronic
spectra, and in the absence of nonadiabatic couplings. Indeed, the
time-resolved spectroscopic signals have been successfully evaluated using the
thawed Gaussian approximation on such model potentials.\cite{Braun_Engel:1998,
Rohrdanz_Cina:2006} To justify our prediction of time-resolved electronic
spectra of the phenyl radical, we construct a Morse potential corresponding to
the single vibrational mode and evaluate the pump-probe spectra using both the
thawed Gaussian approximation and exact quantum dynamics. The results imply
that the thawed Gaussian approximation is rather accurate at this level of
anharmonicity, but also that more-dimensional calculations are required to
reproduce the on-the-fly \emph{ab initio} spectrum.

\section{\label{sec:theory}Theory}

\subsection{\label{subsec:spec}Vibrationally and time-resolved electronic
spectroscopy}

Let us consider a system described by the total Hamiltonian
\begin{equation}
\mathbf{\hat{H}}_{\text{tot}}(t)=\mathbf{\hat{H}}+\mathbf{\hat{V}}%
_{\text{int}}(t) \label{eq:H_tot}%
\end{equation}
where $\mathbf{\hat{H}}$ is the molecular Hamiltonian and $\mathbf{\hat{V}%
}_{\text{int}}(t)$ represents the interaction potential with the external
field $\vec{E}(t)$ within the electric-dipole approximation:
\begin{equation}
\mathbf{\hat{V}}_{\text{int}}(t)=-\bm{\hat{\vec{\mu}}}\cdot\vec{E}(t),
\label{eq:V_int}%
\end{equation}
where $\bm{\hat{\vec{\mu}}}$ stands for the molecular electric dipole
operator. Above, we have introduced notation, in which the \textbf{bold} font
is used for the matrix representation of electronic operators, the nuclear
operators are distinguished by the hat $\,\hat{}\,$ symbol, and the arrows
denote three-dimensional vectors.

In a pump-probe experiment, the electric field consists of two components,
corresponding to the pump and probe pulses:
\begin{equation}
\vec{E}(t):=\vec{\epsilon}^{\text{ pu}}E^{\text{pu}}(t)+\vec{\epsilon}^{\text{
pr}}E^{\text{pr}}(t), \label{eq:E_field}%
\end{equation}
where $\vec{\epsilon}$ is the polarization vector and
\begin{align}
E^{\text{pu}}(t)  &  :=E_{e}^{\text{pu}}(t)\delta_{n}(t)\,,\label{eq:E_pu}\\
E^{\text{pr}}(t)  &  :=E_{e}^{\text{pr}}(t)\delta_{n}(t-\tau)\,.
\label{eq:E_pr}%
\end{align}
Here, $E_{e}^{\text{pu}}(t)$ and $E_{e}^{\text{pr}}(t)$ represent the rapidly
changing factors of the pump and probe electric fields (i.e., the factors that
oscillates on the electronic time scale), whereas $\delta_{n}$ is a slowly
varying envelope; the pump pulse is centered at time zero and the probe pulse
is centered at a delay time $\tau$. For simplicity, we assume that both
envelopes are the same and that the fields are linearly polarized. Therefore,
we can first keep the molecular orientation fixed and consider only the
projections of the dipole operator onto the polarizations of the fields:
\begin{align}
\bm{\hat{\mu}}^{\text{pu}}  &  :=\bm{\hat{\vec{\mu}}}\cdot\vec{\epsilon
}^{\text{ pu}},\label{eq:mu_pu}\\
\bm{\hat{\mu}}^{\text{pr}}  &  :=\bm{\hat{\vec{\mu}}}\cdot\vec{\epsilon
}^{\text{ pr}}, \label{eq:mu_pr}%
\end{align}
keeping in mind that in an isotropic sample, in the end one has to average
over all possible orientations of the molecule with respect to the external
electric fields.

\subsection{\label{subsec:pert_rho}Perturbative expressions for the
time-resolved electronic spectra}

Differential absorption cross-section, which is typically \emph{the}
time-resolved spectrum\ measured in the pump-probe experiment, is the
difference between the pump-on and pump-off spectra of the probe pulse
\begin{equation}
\sigma(\omega,\tau)=\sigma^{\text{pu}+\text{pr}}(\omega,\tau)-\sigma
^{\text{pr}}(\omega). \label{eq:sigma_PP_general}%
\end{equation}

Numerous approaches can be taken to evaluate the differential absorption
spectrum, including both perturbative and non-perturbative
methods.\cite{Seidner_Domcke:1992,Mukamel:1995} The perturbative
approach\cite{Mukamel:1995, Domcke_Stock:1997,
Pollard_Mathies:1990a,Tannor:2007,Cina_Scholes:2016} requires the third-order
time-dependent perturbation theory; we employ this approach here, adopting the
notation and derivation of Ref.~\onlinecite{Zimmermann_Vanicek:2014}. While
higher-order perturbative expressions are simpler for the density operators,
the thawed Gaussian approximation relies on the wavefunction representation.
The connection between the density-matrix and wavefunction formalisms is
described, e.g., in the monographs by Mukamel\cite{Mukamel:1995} and
Tannor;\cite{Tannor:2007} therefore, in the main text we only mention the
relevant approximations, deferring the supporting equations to the Appendix.

Several approximations simplify the analysis of pump-probe experiments substantially:

\begin{enumerate}
\item Nonoverlapping pulses---all interactions with the probe pulse follow all
interactions with the pump pulse.

\item Ultrashort pulse (or impulsive) approximation---the pulse envelopes
$\delta_{n}$ are assumed to be delta functions on the nuclear time scale, but
long on the electronic time scale; thus, we assume that the wavepacket is
created only in one electronic state.

\item Resonance condition---laser fields are almost in resonance with the
electronic transitions.

\item Phase matching---overall momentum transfer between the molecule and the
pump pulse is approximately zero, i.e., $\hbar\vec{k}_{1}^{\text{pu}}%
-\hbar\vec{k}_{2}^{\text{pu}}\approx0$ because we are interested in the signal
in the direction of the probe pulse.
\end{enumerate}

Within these approximations, the third-order time-dependent perturbation
theory expression for the pump-probe spectrum
is\cite{Zimmermann_Vanicek:2014}
\begin{equation}
\sigma^{(3)}(\omega,\tau)=\frac{4\pi\omega}{\hbar c}\text{Re}\int_{0}^{\infty
}\left[  C_{\mu,\tau}^{(3)}(t^{\prime})^{\ast}-C_{\mu,\tau}^{(3)}(t^{\prime
})\right]  e^{i\omega t^{\prime}}dt^{\prime}, \label{eq:sigma_PP}%
\end{equation}
where $t^{\prime}$ is the time after the probe pulse, i.e., $t^{\prime
}:=t-\tau$ and $C_{\mu,\tau}^{(3)}$ is the third-order dipole time
autocorrelation
\begin{equation}
C_{\mu,\tau}^{(3)}(t^{\prime})=\text{Tr}[\bm{\hat{\rho}}^{\text{pu}%
}\bm{\hat{\mu}}^{\text{pr}}(\tau)\bm{\hat{\mu}}^{\text{pr}}(\tau+t^{\prime})],
\label{eq:C_mu_tau}%
\end{equation}
with
\begin{equation}
\bm{\hat{\rho}}^{\text{pu}}=2\pi^{2}\hbar^{-2}[\bm{\hat{\mu}}_{\text{RC}%
}^{\text{pu}},[\bm{\hat{\rho}},\bm{\hat{\mu}}_{\text{RC}}^{\text{pu}}]]
\label{eq:rho_pu}%
\end{equation}
and the molecular dipole operator in the Heisenberg picture
\begin{equation}
\bm{\hat{\mu}}^{\text{pr}}(t)=e^{i\mathbf{\hat{H}}(t)/\hbar}%
\bm{\hat{\mu}}^{\text{pr}}e^{-i\mathbf{\hat{H}}(t)/\hbar}. \label{eq:mu_pr_t}%
\end{equation}
In Eq.~(\ref{eq:rho_pu}), $\bm{\hat{\rho}}$ is the initial (stationary)
density and $\bm{\hat{\mu}}_{\text{RC}}^{\text{pu}}$ is the electric dipole
operator which incorporates the resonant condition, i.e., its matrix elements
corresponding to the transitions between the electronic states $j$ and $k$ are
scaled by the Fourier transform of the pump electric field evaluated at the
transition frequency $\omega_{jk}$ between those electronic states:
\begin{equation}
\bm{\hat{\mu}}_{\text{RC}}^{\text{pu}}:=\sum_{jk}\tilde{E}^{\text{pu}}%
(\omega_{jk})\hat{\mu}_{jk}^{\text{pu}}\left\vert j\right\rangle \left\langle
k\right\vert .
\end{equation}
Equation~(\ref{eq:C_mu_tau}) for $C_{\mu,\tau}^{(3)}(t^{\prime})$\ is expanded
using Eqs.~(\ref{eq:rho_pu})--(\ref{eq:mu_pr_t}) in Appendix
\ref{sec_app:C_mu_tau}.

To derive the final and more specific expressions for the pump-probe spectra
used in this paper, we also assume:

\begin{enumerate}
\item Three-state system---only the ground and two excited electronic states
are relevant for the description of the whole system, whereas other states are
either dark or not in resonance; the three states are approximately equally
spaced, so that the electric fields are in resonance with both 0--1 and 1--2
transitions, but not with the direct 0--2 electronic transition.

\item Zero-temperature limit---only the ground vibrational state of the ground
electronic state is initially populated, i.e., $\hat{\rho}_{00}=\left\vert
\psi_{0,g}\left\rangle {}\right\langle \psi_{0,g}\right\vert $, and all other
density matrix elements are zero.

\item Born--Oppenheimer approximation---there is no population transfer
between electronic states during the evolution with the molecular Hamiltonian.

\item Condon approximation---the electronic transition dipole moment matrix
elements are independent of nuclear coordinates.
\end{enumerate}

The first approximation is an expanded form of the usual two-state model used
for the description of linear absorption and emission spectra; here, a third
state is needed to account for the excited-state absorption. (Of course, in
practice, there may be many more relevant excited states.) Our molecular
Hamiltonian is a diagonal $3\times3$ matrix in electronic states with nuclear
Hamiltonians $\hat{H}_{0}$, $\hat{H}_{1}$, and $\hat{H}_{2}$ on the diagonal.
As for the zero-temperature limit, the assumption that only the ground
electronic state is populated is, of course, satisfied to much higher accuracy
than the assumption that only the ground vibrational level is occupied.
Nevertheless, the main contributions to the spectrum are often captured even
within the zero-temperature limit. As for the validity of the
Born--Oppenheimer approximation, the nonadiabatic effects can introduce a
decay of the spectrum intensity with increasing probe delay time $\tau
$.\cite{Stock_Domcke:1993,Zimmermann_Vanicek:2014} For the short times
discussed in this work, such decay can be neglected or subsequently added
phenomenologically. However, as pointed out in
Ref.~\onlinecite{Stock_Domcke:1993}, this does not hold for longer probe delay
times. Finally, we note that the Condon approximation is used only at the last
step of the derivation; similar expressions are obtained even beyond the
Condon approximation.

As shown in Appendix \ref{sec_app:sigma_PP_CA_final}, the main contributions
to the absorption spectrum of Eq.~(\ref{eq:sigma_PP}) are the excited-state
absorption, stimulated emission, and ground-state bleach:
\begin{align}
\sigma^{(3)}(\omega,\tau)  &  =\frac{16\pi^{3}\omega|\tilde{E}^{\text{pu}%
}(\omega_{10})|^{2}}{\hbar^{3}c}\left\vert \mu_{10}^{\text{pu}}\right\vert
^{2}\nonumber\\*
\times &  \text{Re}\int_{0}^{\infty}\left[  \left\vert \mu_{21}^{\text{pr}%
}\right\vert ^{2}C_{\text{ESA}}^{\ast}(t^{\prime},\tau)-\left\vert \mu
_{10}^{\text{pr}}\right\vert ^{2}C_{\text{SE}}(t^{\prime},\tau)-\left\vert
\mu_{10}^{\text{pr}}\right\vert ^{2}C_{\text{GSB}}(t^{\prime},\tau)\right]
e^{i\omega t^{\prime}}dt^{\prime},\label{eq:sigma_PP_CA_final}\\
C_{\text{ESA}}(t^{\prime},\tau)  &  :=\left\langle \psi_{0,g}\left\vert
e^{i\hat{H}_{1}\tau/\hbar}e^{i\hat{H}_{2}t^{\prime}/\hbar}e^{-i\hat{H}%
_{1}(t^{\prime}+\tau)/\hbar}\right\vert \psi_{0,g}\right\rangle
,\label{eq:ESA}\\
C_{\text{SE}}(t^{\prime},\tau)  &  :=\left\langle \psi_{0,g}\left\vert
e^{i\hat{H}_{1}\tau/\hbar}e^{i\hat{H}_{0}t^{\prime}/\hbar}e^{-i\hat{H}%
_{1}(t^{\prime}+\tau)/\hbar}\right\vert \psi_{0,g}\right\rangle ,
\label{eq:SE}\\
C_{\text{GSB}}(t^{\prime},\tau)  &  :=\left\langle \psi_{0,g}\left\vert
e^{i\hat{H}_{0}t^{\prime}/\hbar}e^{-i\hat{H}_{1}t^{\prime}/\hbar}\right\vert
\psi_{0,g}\right\rangle . \label{eq:GSB}%
\end{align}
The ground-state bleach is, within the given approximations, equal to the
scaled continuous-wave absorption spectrum. If the ultrashort pulse
approximation is relaxed, the ground-state bleach is better described as
(impulsive) stimulated resonant Raman
scattering\cite{Pollard_Mathies:1990a,Pollard_Mathies:1992} because the
short-time dynamics on the excited state, which can proceed between the two
interactions with the pump pulse, creates a wavepacket on the ground
electronic state. However, the dynamics of such wavepacket is limited because
it does not have time to move far away from the Franck--Condon region during
the short pump pulse. In contrast, the stimulated emission and excited-state
absorption signals can drift significantly from their original positions
because the excited-state dynamics occurs during the delay time $\tau$.
Consequently, the effect of a finite pump pulse can be safely neglected.

Finally, the spectrum of Eq.~(\ref{eq:sigma_PP_CA_final}) must be averaged
over all possible orientations of the molecule with respect to the electric
fields. Within the Condon approximation, the averaging results only in a
constant scaling factor, which depends on the angle between the polarizations
of the pump and probe pulses, ranging from $1/15$ for perpendicular to $1/5$
for parallel polarizations (see Appendix \ref{sec_app:orient_av}).

\subsection{\label{subsec:trse}Wavepacket approach to time-resolved stimulated
emission}

In the following, we concentrate on the time-resolved stimulated emission
spectrum and only in the final result we add the ground-state bleach; the
excited-state absorption signal can be computed in a similar way as the
stimulated emission. Although the excited-state absorption is not negligible,
it is commonly found at different frequencies from the other two contributions
and can be analyzed separately, unlike the ground-state bleach and stimulated
emission, which often overlap.\cite{Berera_Kennis:2009}

The stimulated emission contribution to the correlation function can be
rewritten in terms of a correlation
function,\cite{Pollard_Mathies:1990a,Mukamel:1995} sometimes referred to as
fidelity amplitude,\cite{Gorin_Znidaric:2006,Vanicek:2017}
\begin{equation}
C_{\text{SE}}(t^{\prime},\tau)=\left\langle \psi_{\tau}\left\vert e^{i\hat
{H}_{0}t^{\prime}/\hbar}e^{-i\hat{H}_{1}t^{\prime}/\hbar}\right\vert
\psi_{\tau}\right\rangle \label{eq:SE_fidelity}%
\end{equation}
by defining a non-stationary wavepacket
\begin{equation}
|\psi_{\tau}\rangle:=e^{-i\hat{H}_{1}\tau/\hbar}|\psi_{0,g}\rangle\,.
\label{eq:psi_tau}%
\end{equation}
To evaluate Eq.~(\ref{eq:SE_fidelity}), one must first propagate the nuclear
wavepacket on the excited-state surface for time $\tau$ [according to
Eq.~(\ref{eq:psi_tau})], and then propagate it simultaneously on both ground-
and excited-state surfaces for time $t^{\prime}$. Because the nonstationary
state $\psi_{\tau}$ depends on $\tau$, every choice of the probe delay time
$\tau$ requires a new full $t^{\prime}$ propagation on the two surfaces and
every $t^{\prime}$ propagation must be long enough to obtain vibrational
resolution. In contrast, one or only several delays have to be considered,
since the time $\tau$ is not related to the vibrational resolution of the spectra.

\subsection{\label{subsec:tga}On-the-fly \emph{ab initio} thawed Gaussian
approximation}

The thawed Gaussian approximation is one of the simplest approaches to
evolving a wavepacket in a general potential. It considers a Gaussian
wavepacket, described by its time-dependent position $q_{t}$, momentum $p_{t}%
$, complex symmetric matrix $A_{t}$, and a complex number $\gamma_{t}$,
\begin{equation}
\psi(q,t)=N_{0}\exp\left\{  -(q-q_{t})^{T}\cdot A_{t}\cdot(q-q_{t})+\frac
{i}{\hbar}\left[  p_{t}^{T}\cdot(q-q_{t})+\gamma_{t}\right]  \right\}  .
\end{equation}
In a harmonic potential, which can even be time-dependent, the Gaussian form
of the wavepacket is exactly preserved---the propagation of the wavepacket
only requires evaluating the four time-dependent parameters $A_{t}$, $q_{t}$,
$p_{t}$, and $\gamma_{t}$.\cite{Heller:1975,Coalson_Karplus:1990} This forms
the basis of the thawed Gaussian approximation, where the potential is given
by the local harmonic approximation
\begin{equation}
V_{\text{LHA}}(q,t)=V(q_{t})+\text{grad}_{q}V|_{q_{t}}\cdot(q-q_{t})+\frac
{1}{2}(q-q_{t})^{T}\cdot\text{Hess}_{q}V|_{q_{t}}\cdot(q-q_{t})
\end{equation}
of the exact potential $V(q)$ about the center $q_{t}$ of the wavepacket at
time $t$. The equations of motion for the time-dependent parameters
are\cite{Heller:1975}
\begin{align}
\dot{q}_{t}  &  =m^{-1}\cdot p_{t}\,,\\
\dot{p}_{t}  &  =-\text{grad}_{q}\,V|_{q_{t}}\,,\\
\dot{A}_{t}  &  =-2i\hbar A_{t}\cdot m^{-1}\cdot A_{t}+\frac{i}{2\hbar
}\text{Hess}_{q}V|_{q_{t}}\,,\\
\dot{\gamma}_{t}  &  =L_{t}-\hbar^{2}\text{Tr}\left(  m^{-1}\cdot
A_{t}\right)  \,,
\end{align}
where $m$ is the mass matrix and $L_{t}$ the Lagrangian. Since the thawed
Gaussian approximation requires only a single classical trajectory, along with
the corresponding Hessians, it is suitable for an on-the-fly implementation,
where the potential information is obtained as-needed from an \emph{ab initio}
electronic structure code. The details of our on-the-fly implementation and,
in particular, the conversion from the Cartesian to normal mode coordinates,
can be found in Refs.~\onlinecite{Wehrle_Vanicek:2014, Wehrle_Vanicek:2015}.

The local harmonic approximation holds as long as the wavepacket is
compact---the spreading of the wavepacket leads to incorrect description of
the potential at the tails. Consequently, the single Gaussian wavepacket
ansatz cannot describe wavepacket splitting or tunneling through barriers.
Nevertheless, the thawed Gaussian approximation has two extremely important
properties for molecular quantum dynamics and, in particular, evaluation of
vibrationally resolved electronic spectra. First, it is exact in a global
harmonic potential, which is often a decent first approximation to the exact
potential energy surface, and performs well for harmonic-like potentials,
outperforming the global harmonic approaches due to the partial inclusion of
anharmonicity. Second, it is valid for short times before the wavepacket
splitting takes effect, which is useful for describing low-resolution vibronic
spectra and ultrafast time-resolved experiments.

The computational cost of the Hessian evaluation is typically significantly
higher than the cost of the corresponding gradients. To address this issue, we
(i) evaluate some of the Hessian values in parallel and (ii) interpolate the
other Hessian values subsequently. The Hessian of the potential is not needed
for the classical trajectory propagation, which allows us to first propagate
the classical trajectory and then evaluate the Hessians at different points
along the trajectory in parallel. In addition, the Hessian of the potential
changes more slowly than does the gradient, allowing us to evaluate the
Hessian only every several steps, and interpolate in between. These two tricks
reduce the computational time significantly, with almost no effect on the
accuracy. Note that several alternative approaches, based on Hessian
updates\cite{Ceotto_Hase:2013,Zhuang_Ceotto:2013} or Gaussian process
regression,\cite{Alborzpour_Habershon:2016, Laude_Richardson:2018} can be used
as well.

\section{\label{sec:compdet}Computational details}

For the electronic structure calculations, i.e., the evaluation of the
energies, gradients, and Hessians, we used density functional theory for the
ground state and time-dependent density functional theory for the excited
electronic state, as implemented in the Gaussian09\cite{g09} electronic
structure package, with B3LYP functional and SNSD basis
set;\cite{Baiardi_Barone:2013} this \emph{ab initio} method was validated in a
previous work.\cite{Patoz_Vanicek:2018}

\begin{figure}
[ptb]%
\includegraphics[scale=1]{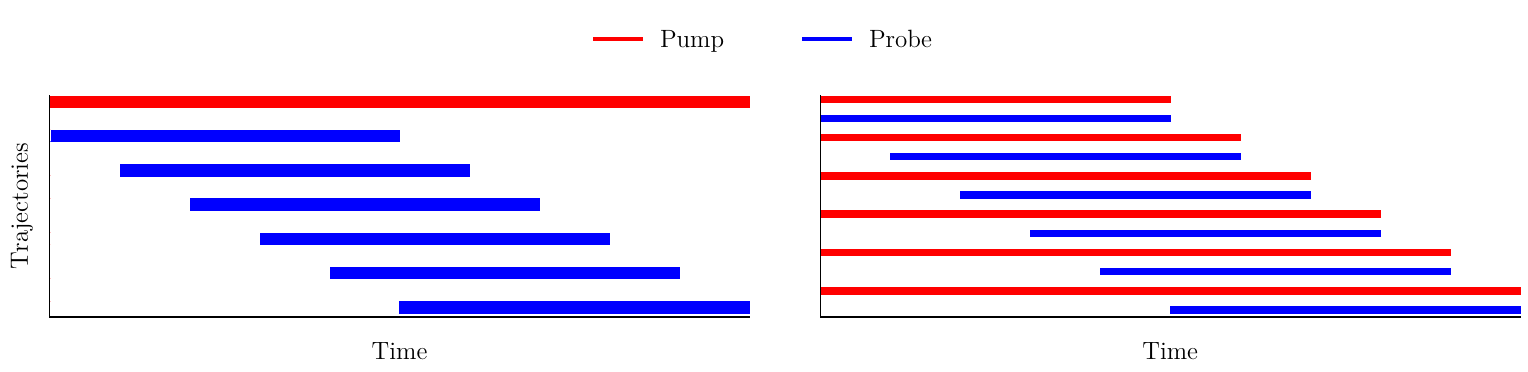}\caption{We first evaluate a
long ``pump trajectory'' moving on the excited-state surface and then use its snapshots as initial conditions
for ``probe trajectories'' moving on the ground-state surface and for evaluating the correlation functions (left
panel), instead of recomputing the pump trajectory for each probe delay time
(right panel).}\label{PP_Schematic}
\end{figure}

The time step was 8 a.u. ($\approx0.194$ fs). The time-resolved spectra were
computed for 65 equidistant probe delays with a time separation of 12 steps
($\approx2.3$ fs), resulting in a total time of about $150$ fs. The
corresponding correlation functions were computed according to
Eq.~(\ref{eq:SE_fidelity}); the wavepackets were propagated for 1500 steps
($\approx290$ fs) on the ground- and excited-state surfaces. The \emph{ab
initio} data needed for the wavepacket propagation were acquired in the
following way (see Fig.~\ref{PP_Schematic}): First, a longer \textquotedblleft
pump\textquotedblright\ classical trajectory was propagated on the excited
electronic state to cover all probe times, i.e., for time $150+290=440$ fs.
Then, 65 probe trajectories were run, each with its own initial conditions
taken from the snapshots of the pump trajectory at the probe delay times.
Finally, the Hessians were evaluated in parallel for all trajectories every 8
steps of the calculation. The described procedure avoids recomputing the same
part of the pump trajectory. Once the \emph{ab initio} data were collected,
the wavepacket was propagated in the mass-scaled normal mode coordinates of
the ground electronic state. The initial wavepacket was taken as the ground
vibrational state of the electronic ground-state potential energy surface
approximated by a harmonic potential, i.e., a Gaussian centered at the
ground-state minimum geometry with a width determined by the Hessian of the
electronic ground-state surface.

The population decay and the interactions with the environment were neglected
during the short-time excited-state dynamics, while the intramolecular nuclear
dynamics was fully accounted for with our full-dimensional wavepacket
propagation method. We applied a phenomenological inhomogeneous Gaussian
broadening (half-width at half-maximum of 100 cm$^{-1}$) of the spectra by
multiplying the correlation functions by a Gaussian decay function. Since no
particular choice of the pump field is taken, the absolute scaling of the
spectrum is not well defined---all spectra were rescaled as indicated in the
corresponding figure captions.

\section{\label{sec:resanddisc}Results and discussion}

The time-resolved electronic spectrum of phenyl radical is evaluated as a sum
of the stimulated emission and ground-state bleach contributions; the
excited-state absorption term is not included. The stimulated emission term
determines the overall dynamics of the spectra, whereas the ground-state
bleach is, within our assumptions, independent of time and proportional to the
scaled linear absorption spectrum. We compare three methods for evaluating the
spectra: the first employs the on-the-fly \emph{ab initio} thawed Gaussian
approximation, which includes the anharmonicities of both ground- and
excited-state surfaces, the second---global adiabatic harmonic
method---expands the two potential energy surfaces to the second order about
the corresponding minima, whereas the third method combines the on-the-fly
approach for the excited state with the global adiabatic harmonic approach for
the ground state.

To analyze a calculated spectrum in more detail, we evaluate the mean (i.e.,
the first moment)
\begin{equation}
\langle\omega\rangle_{\tau}:=\int\omega\sigma_{0}(\omega,\tau)d\omega
\label{eq:mean}%
\end{equation}
and width, measured by the standard deviation (i.e., the second central
moment)
\begin{equation}
\Delta\omega_{\tau}:=\sqrt{\langle\omega^{2}\rangle_{\tau}-\langle
\omega\rangle_{\tau}^{2}}, \label{eq:var}%
\end{equation}
of the normalized spectral
lineshapes\cite{Pollard_Mathies:1990a,Ferrer_Santoro:2013}
\begin{equation}
\sigma_{0}(\omega,\tau)=\frac{\sigma(\omega,\tau)/\omega}{\int(\sigma
(\omega,\tau)/\omega)d\omega} \label{eq:sigma_normalized}%
\end{equation}
as functions of the delay time $\tau$. The two spectral moments are closely
related to the properties of the nuclear wavepacket and its dynamics on the
excited-state surface. The mean is determined by the expectation value of the
difference $\Delta\hat{V}$ between the excited and ground potential
energies:\cite{Pollard_Mathies:1990a}
\begin{equation}
\langle\omega\rangle_{\tau}=\frac{1}{\hbar}\langle\psi_{\tau}|\Delta\hat
{V}|\psi_{\tau}\rangle, \label{eq:mean_wp}%
\end{equation}
whereas the standard deviation of the spectrum is given by the standard
deviation of $\Delta\hat{V}$:\cite{Pollard_Mathies:1990a}
\begin{equation}
\Delta\omega_{\tau}=\frac{1}{\hbar}\sqrt{\langle\psi_{\tau}|\Delta\hat{V}%
^{2}|\psi_{\tau}\rangle-\langle\psi_{\tau}|\Delta\hat{V}|\psi_{\tau}%
\rangle^{2}} \label{eq:var_wp}%
\end{equation}
at time $\tau$.

To simplify the expressions for the mean and width of spectra, let us
introduce the shorthand notation
\begin{equation}
\Sigma_{\tau}^{2}:=\frac{1}{4}(\operatorname{Re}A_{\tau})^{-1}
\label{eq:Gamma}%
\end{equation}
for the inverse of the real part of the width matrix, and
\begin{align}
\Delta V_{\tau}  &  :=\Delta V|_{q_{\tau}},\label{eq:delta_V_tau}\\
\Delta V^{\prime}_{\tau}  &  :=\Delta\operatorname{grad}_{q}V|_{q_{\tau}%
},\label{eq:delta_F_tau}\\
\Delta V^{\prime\prime}_{\tau}  &  :=\Delta\operatorname{Hess}_{q}V|_{q_{\tau
}}. \label{eq:delta_H_tau}%
\end{align}
for the differences of energies, forces, and Hessians of the two potential
energy surfaces at the current position of the wavepacket. Then, within the
local harmonic approximation for the potential and assuming the thawed
Gaussian wavepacket, one can analytically evaluate Eq.~(\ref{eq:mean_wp}) for
the spectral mean, to obtain
\begin{equation}
\langle\omega\rangle_{\tau}^{\text{LHA}}=\frac{1}{\hbar}\left[  \Delta
V_{\tau}+\frac{1}{2}\text{Tr}(\Delta V^{\prime\prime}_{\tau} \cdot\Sigma
_{\tau}^{2})\right]  , \label{eq:mean_LHA}%
\end{equation}
and Eq.~(\ref{eq:var_wp}) for the spectral width, to obtain
\begin{align}
\Delta\omega_{\tau}^{\text{LHA}}  &  =\frac{1}{\hbar}\Big[(\Delta V^{\prime
}_{\tau})^{T}\cdot\Sigma_{\tau}^{2}\cdot\Delta V^{\prime}_{\tau}\nonumber\\
&  +\frac{1}{2}\text{Tr}(\Delta V^{\prime\prime}_{\tau} \cdot\Sigma_{\tau}^{2}
\cdot\Delta V^{\prime\prime}_{\tau} \cdot\Sigma_{\tau}^{2})\Big]^{\frac{1}{2}%
}. \label{eq:var_LHA}%
\end{align}
If the two potentials have similar curvatures, i.e. $\Delta V^{\prime\prime
}_{\tau}\approx0$, the mean of the spectrum corresponds to the energy gap
evaluated at the center of the Gaussian wavepacket, i.e.,
\begin{equation}
\left\langle \omega\right\rangle _{\tau}^{\text{approx.}}=\Delta V_{\tau
}/\hbar, \label{eq:mean_approx}%
\end{equation}
while Eq.~(\ref{eq:var_LHA}) for the width reduces to
\begin{equation}
\Delta\omega_{\tau}^{\text{approx.}}=\frac{1}{\hbar}\sqrt{(\Delta V^{\prime
}_{\tau})^{T}\cdot\Sigma_{\tau}^{2}\cdot\Delta V^{\prime}_{\tau}},
\label{eq:var_approx}%
\end{equation}
showing that the width of the spectrum depends both on the width of the
nuclear wavepacket and the gradient of the energy gap evaluated at the center
of the wavepacket. Equations~(\ref{eq:mean_LHA}) and (\ref{eq:var_LHA}) are
exact for general harmonic potentials, while Eqs.~(\ref{eq:mean_approx}) and
(\ref{eq:var_approx}) are exact for displaced harmonic oscillators with the
same force constants. In addition, the evaluation of the four expressions for
the mean and width does not require running any probe trajectories; all that
is needed is the excited-state thawed Gaussian propagation and the additional
information on the ground-state potential evaluated along the excited-state trajectory.

Finally, let us point out that the broadening of spectra introduced as a decay
of the correlation function can affect the width of the spectra in a nonlinear
way. In particular, for the Gaussian broadening of the form $\exp(-\alpha
t^{2})$, equations~(\ref{eq:var_LHA}) and (\ref{eq:var_approx}) for the widths
must be modified to
\begin{equation}
\Delta\omega_{\tau}^{k,\alpha}=\frac{1}{\hbar}\sqrt{(\Delta\omega_{\tau}%
^{k})^{2}+2\alpha}, \label{eq:var_alpha}%
\end{equation}
where $k$ stands for either \textquotedblleft$\text{LHA\textquotedblright\ or
\textquotedblleft approx.\textquotedblright}$, depending on the level of
approximation. Equation~(\ref{eq:var_alpha}) is used in the calculations
presented below. See Appendices \ref{sec_app:position_width} and
\ref{sec_app:position_width_broadening} for the derivations of the expressions
for $\left\langle \omega\right\rangle _{\tau}$ and $\Delta\omega_{\tau}$ given here.

\subsection{On-the-fly \emph{ab initio} time-resolved electronic spectrum}

\begin{figure}
[ptb]\includegraphics[scale=0.7]{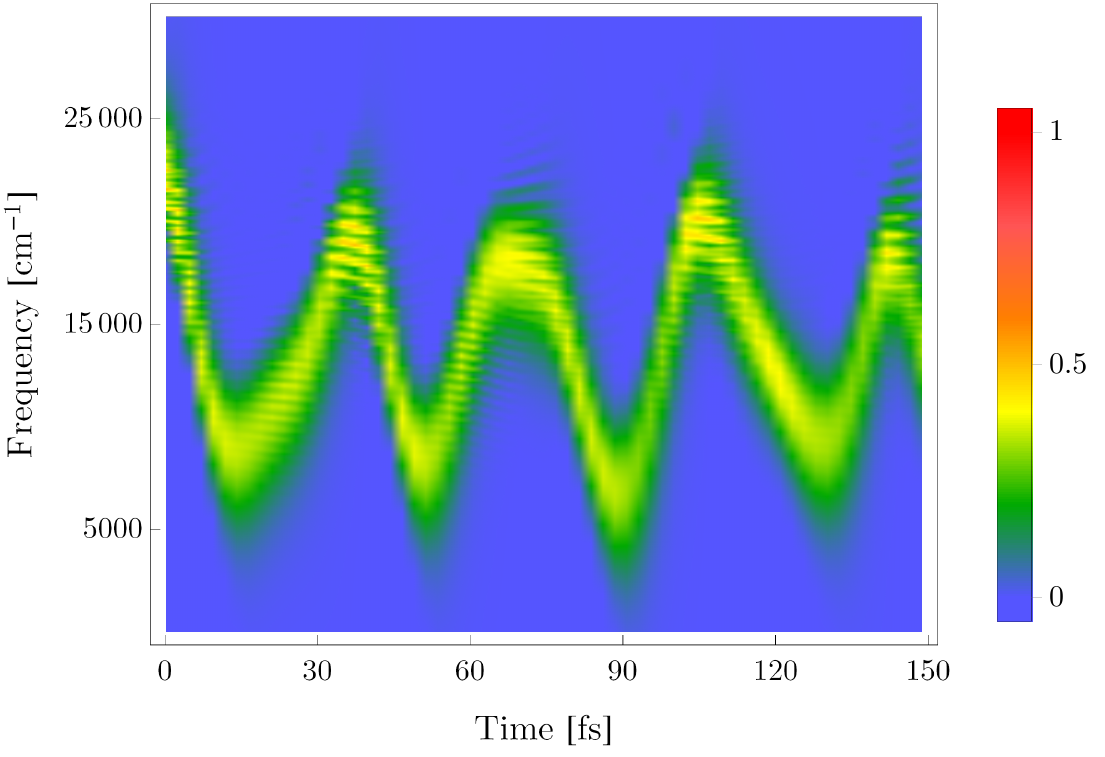}
\includegraphics[scale=0.7]{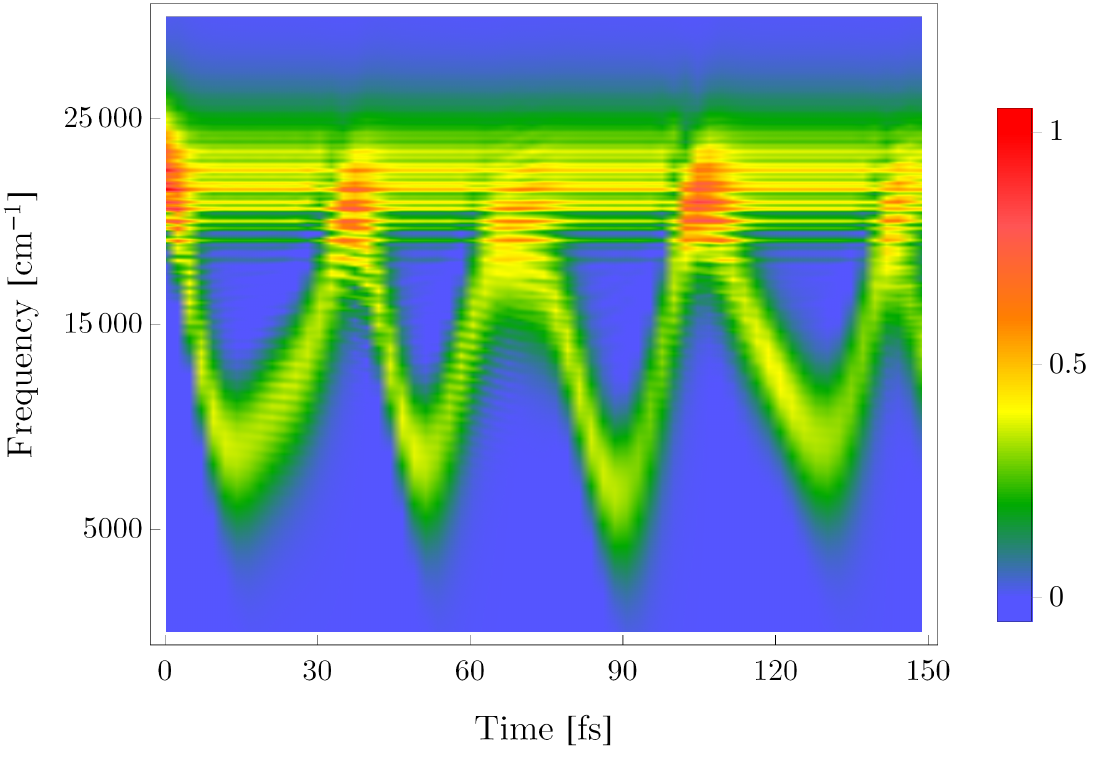}\caption{Time-resolved
stimulated emission spectrum (left) and the pump-probe spectrum including both
the stimulated emission and ground-state bleach (right) of phenyl radical
evaluated with the on-the-fly \emph{ab initio} thawed Gaussian approximation.
Both spectra were rescaled according to the maximum of the time-resolved
spectrum in the right panel.}\label{fig:OTF_TRSE_2D}
\end{figure}

The time-resolved stimulated emission spectrum (Fig.~\ref{fig:OTF_TRSE_2D})
shows oscillations over a broad range of frequencies---a 20000 cm$^{-1}$
window has to be probed to capture all the features. Although the absorption
spectrum covers only about 5000 cm$^{-1}$, the mean of the time-resolved
stimulated emission spectrum moves according to the energy gap between the two
electronic states [Eq.~(\ref{eq:mean_wp})], thus covering different frequency
regions. These oscillations reflect directly the dynamics of the wavepacket on
the excited-state surface: As the wavepacket leaves the Franck--Condon region,
the energy gap decreases, shifting the spectrum towards lower frequencies.
Along with the oscillations in the mean frequency of the spectrum, the
vibrational resolution also changes periodically as a function of the delay
time. In particular, as the wavepacket moves away from the Franck-Condon
region, the spectra become broader and less resolved. The period of
$\approx36$ fs observed in the time-resolved spectrum corresponds to the
frequency of the most-displaced mode in the excited electronic state (924
cm$^{-1}$). Similar observations, namely the high-amplitude oscillations of
the time-resolved stimulated emission spectra and nearly constant ground-state
bleach signal, have been found in calculations based on the harmonic
models\cite{Pollard_Mathies:1990a,Pollard_Mathies:1992} as well as in
experiments.\cite{Fragnito_Shank:1989,Veen_Chergui:2011}

\begin{figure}
[ptb]\includegraphics[scale=1]{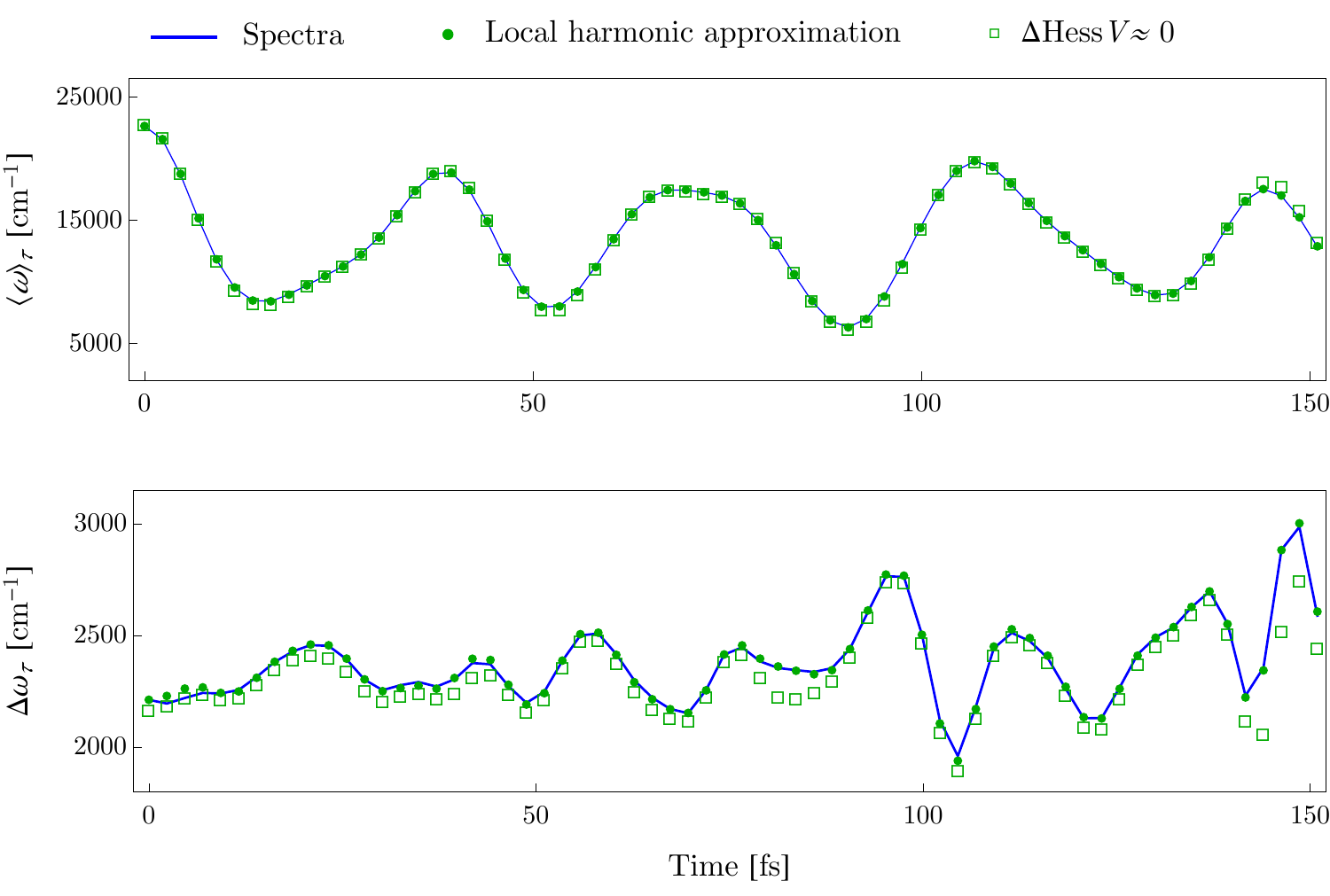}\caption{Means (top panel)
and widths (measured by the standard deviation, bottom panel) of the time-resolved stimulated emission spectra: blue
lines are computed from the spectra, using the general definitions of the mean
[Eq.~(\ref{eq:mean})] and standard deviation [Eq.~(\ref{eq:var})], the green
dots are evaluated from the semiclassical trajectory within the local harmonic approximation
[Eq.~(\ref{eq:mean_LHA}) and Eq.~(\ref{eq:var_alpha}) with $k \equiv \text{``LHA''}$], and the green empty
squares are evaluated from the semiclassical trajectory using
Eq.~(\ref{eq:mean_approx}) for the means and Eq.~(\ref{eq:var_alpha}) with $k \equiv \text{``approx.''}$ for the
widths.}\label{fig:PositionsAndWidths}
\end{figure}

The means and widths of the spectra can be reproduced directly from the
parameters of the wavepacket and the difference between the ground- and
excited-state potentials. Figure~\ref{fig:PositionsAndWidths} shows that the
means and widths of the spectra are almost perfectly reproduced within the
local harmonic approximation, while slightly larger errors are observed with
further approximating $\Delta\text{Hess}_{q}V\approx0$. Since the spectra
themselves are computed within the local harmonic approximation, the means and
widths evaluated from wavepacket properties using Eqs.~(\ref{eq:mean_LHA}) and
(\ref{eq:var_alpha}) with $k\equiv\text{\textquotedblleft
LHA\textquotedblright}$ should, indeed, match exactly the means and widths
calculated from spectra [using Eqs.~(\ref{eq:mean}) and (\ref{eq:var})];
however, slight numerical errors result in differences of the order of $80$
cm$^{-1}$ for the means and $40$ cm$^{-1}$ for widths. The errors due to
neglecting the difference between the Hessians of the two potential energy
surfaces are an order of magnitude greater ($\approx800$ cm$^{-1}$ for the
means and $\approx300$ cm$^{-1}$ for the widths). Nevertheless, both the means
and widths are well described even within this additional approximation.
Although the two moments do not represent the full, vibrationally resolved
spectra completely, they are very useful for describing low-resolution
electronic spectra or spectra, where the vibrational resolution is lost (e.g.
in solution). Finally, we show that including the difference between the
Hessians of the two potentials modifies the results only slightly; the main
determinant of the time dependence of the spectral means is the energy gap
evaluated along the classical trajectory, while the spectral widths require
only the differences between the gradients of the two surfaces.

\subsection{Comparison with the global harmonic methods}

\begin{figure}
[ptb]\includegraphics[scale=1]{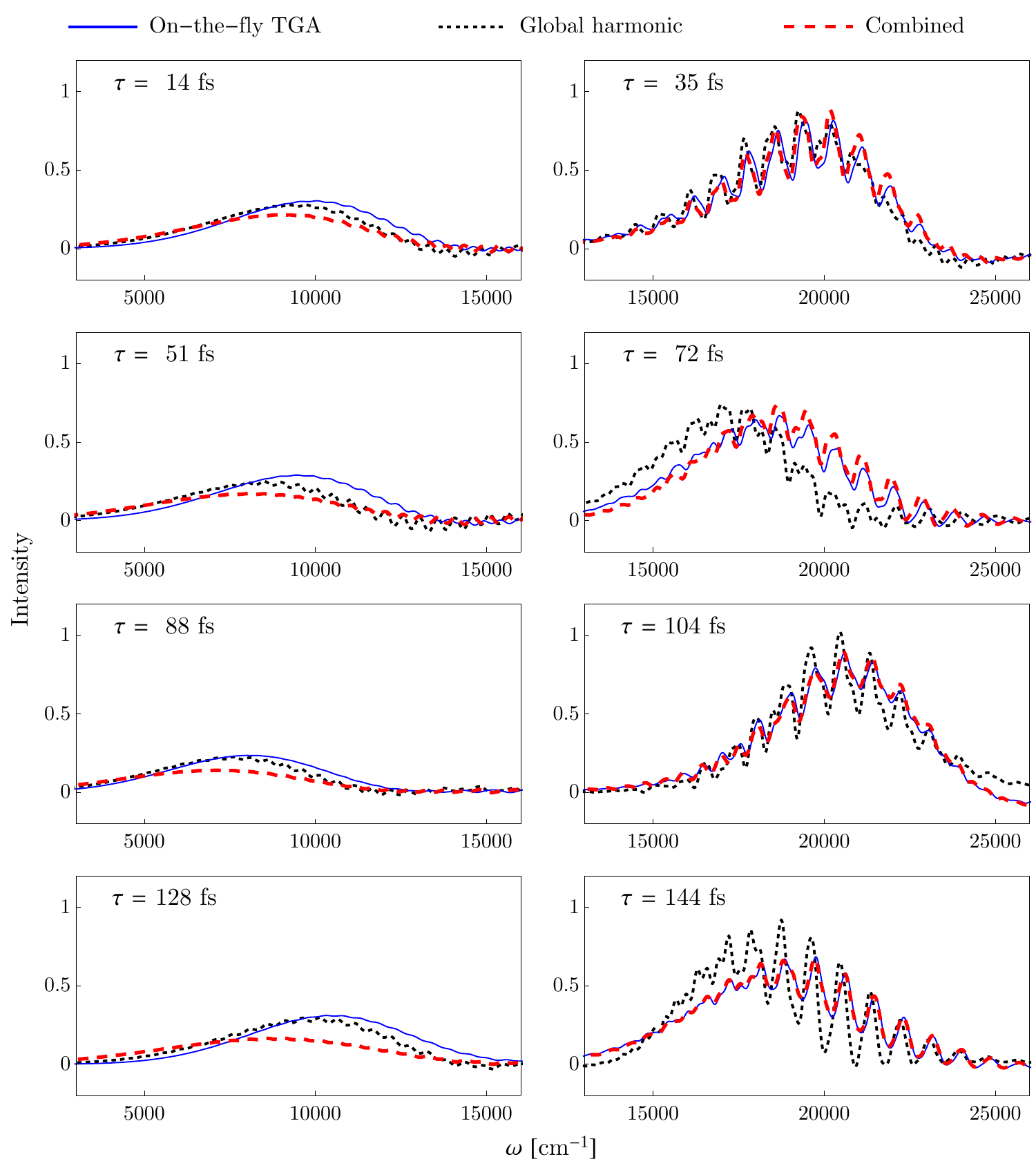}\caption{Time-resolved
stimulated emission spectra of phenyl radical, evaluated using on-the-fly
\emph{ab initio} thawed Gaussian approximation (TGA), global harmonic, and combined
on-the-fly/global harmonic methods, are shown at different delay times $\tau$
of the probe pulse. All spectra are scaled by the maximum of the on-the-fly
spectrum at zero delay. In addition, as done previously in
Ref.~\onlinecite{Patoz_Vanicek:2018}, we introduce frequency shifts by matching the
absorption spectra with the experiment: $-297$ cm$^{-1}$ for the global
harmonic spectrum and $-437$ cm$^{-1}$ for the on-the-fly and combined
on-the-fly/global harmonic spectra.}\label{fig:comparisonplot}
\end{figure}

The on-the-fly approach is the most costly, yet also the most
accurate,\cite{Wehrle_Vanicek:2014,Wehrle_Vanicek:2015,Patoz_Vanicek:2018} of
the three methods discussed here. The least expensive of the three methods and
also the most commonly used approach for evaluating spectra is the adiabatic
harmonic approximation, in which the ground- and excited-state potentials are
expanded to the second order about the corresponding minima. When accurate,
this approximation reduces the computational cost both by avoiding the
evaluation of multiple Hessians and by employing closed-form solutions.
Finally, since the pump-probe spectrum is determined by the dynamics on both
potential energy surfaces, one can also consider combining the on-the-fly and
global harmonic methods. For example, if excited-state dynamics is not
feasible due to the size of the system and/or if the excited-state potential
energy surface is well approximated by a harmonic potential, one can use the
on-the-fly method only for the ground-state dynamics. In contrast, if the
number of the ground-state trajectories is the bottleneck for evaluating the
full time-frequency map, the on-the-fly \emph{ab initio} trajectory may be
computed only for the excited state, and the ground-state surface approximated
by a harmonic potential. This \textquotedblleft combined\textquotedblright%
\ approach is the third of the methods whose merits are analyzed in
Fig.~\ref{fig:comparisonplot}.

\begin{figure}
[ptb]\includegraphics[scale=1]{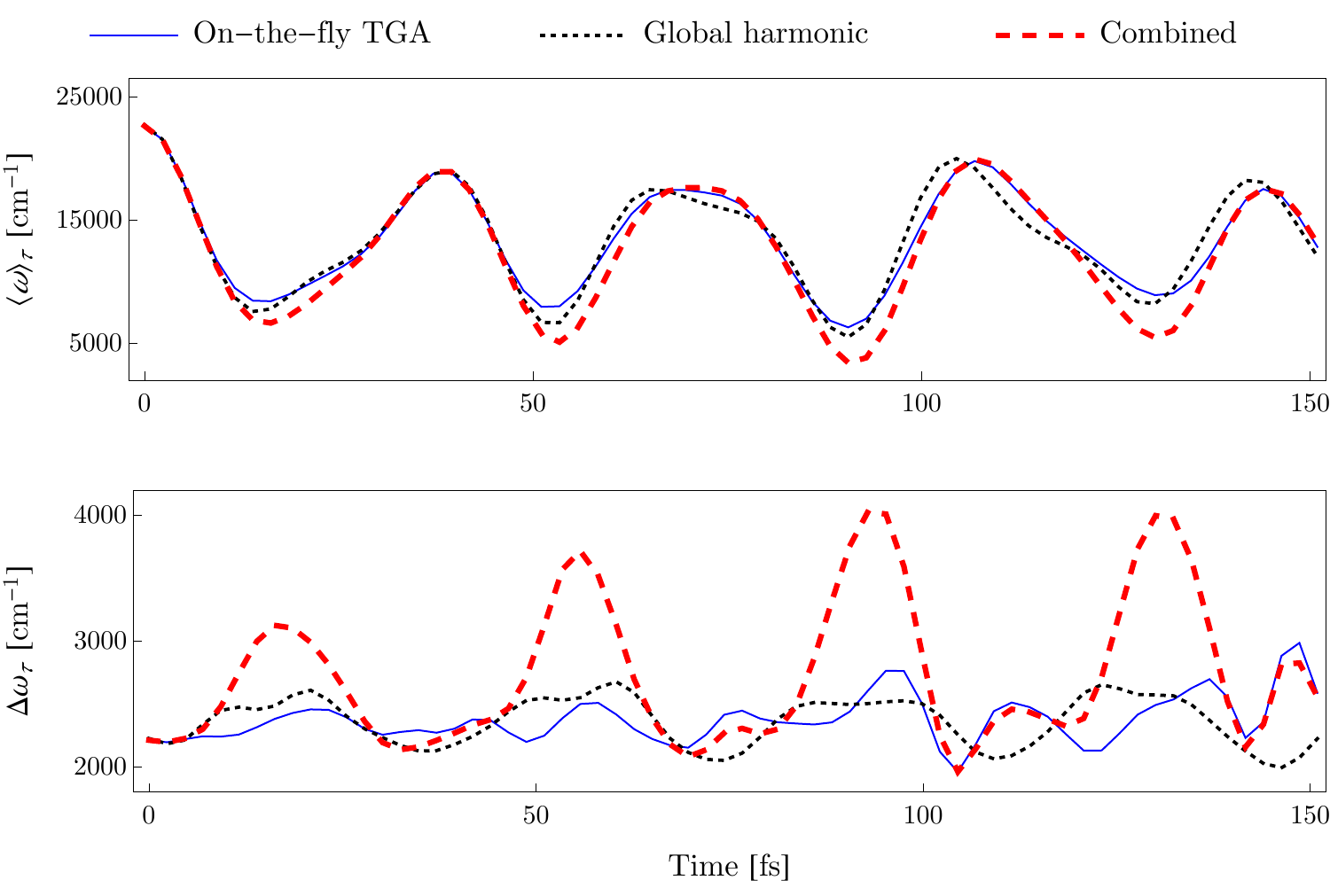}\caption{Means
(top panel) and widths (bottom panel) of the time-resolved spectra evaluated with the
on-the-fly thawed Gaussian approximation (TGA), global harmonic, and combined approaches.}\label{fig:PositionsAndWidthsComparison}%

\end{figure}

The differences observed in Fig.~\ref{fig:comparisonplot} between the global
harmonic and more accurate on-the-fly spectra can be assigned to the
anharmonicity effects, which play a role at all probe delay times. When the
wavepacket is far from the Franck--Condon region, the vibrational resolution
is lost, so the difference can be described by the means and widths of the
spectra (see Fig.~\ref{fig:comparisonplot}, left panels, and
Fig.~\ref{fig:PositionsAndWidthsComparison}). In contrast, when the wavepacket
returns to the initial position, the differences are in the relative
intensities of the individual vibronic bands in the spectrum (see
Fig.~\ref{fig:comparisonplot}, right panels). This pattern is repeated over
time and follows the oscillatory motion of the wavepacket. The errors are,
however, enhanced at later times due to accumulation of the anharmonicity
effects during the dynamics on the excited-state surface. Nevertheless, the
global harmonic approximation offers a decent description of the means and
widths of the spectra, but at a cost of losing the accuracy of the
frequency-resolved features. Interestingly, opposite results are obtained with
the combined on-the-fly/global harmonic approach (see
Fig.~\ref{fig:PositionsAndWidthsComparison}), which fails to reproduce the
on-the-fly means and widths of the spectra at times when the wavepacket is far
from the Franck--Condon region, while describing almost perfectly the
vibrationally resolved spectra when the wavepacket is in the vicinity of the
initial position. This can be understood easily: When the wavepacket returns
to the Franck--Condon region, it is near the ground-state minimum, where the
ground-state potential is well described by the global harmonic approximation.
At the other turning point, far from the Franck--Condon region, the lack of
anharmonicity in the global harmonic approximation of the ground-state
potential leads to the underestimation of the energy gap between the two
surfaces, and, consequently, shifts the combined on-the-fly/global harmonic
spectrum towards lower frequencies. This does not happen when global harmonic
models are applied to both surfaces because the errors due to omitting
anharmonicity in both potential energy surfaces cancel out approximately.

\subsection{Comparison with the exact solution of a one-dimensional model}

Due to the Gaussian ansatz, the thawed Gaussian approximation can break down
at the turning points, where the wavepackets tend to become very asymmetrical
and oscillatory. Such effects are difficult to investigate using steady-state
spectra, which are determined by the autocorrelation function and, therefore,
probe the wavepacket's shape only near the Franck--Condon region.
Time-resolved spectroscopy poses another challenge to wavepacket propagation
methods: when the probe pulse at time $\tau$ sends the nuclear wavepacket from
the excited state back to the ground state, the wavepacket often reaches
regions of higher anharmonicity than in the case of linear, spontaneous or
stimulated emission spectroscopy. To investigate the accuracy of the thawed
Gaussian approximation throughout the wavepacket propagation, we constructed a
simplified one-dimensional model based on the \emph{ab initio} data for the
ground- and excited-state energies of the phenyl radical. The model potential
depends only on the most displaced vibrational mode, neglecting the coupling
to and dynamics in all other modes. Clearly, such a simplified model cannot
describe all spectral features; however, the goal of these calculations is to
provide a similar extent of anharmonicity---which is why we base it on the
\emph{ab initio} data---while allowing for a simple interpretation of the
results. Another motivation for the one-dimensional model is the simple
structure of the pump-probe spectra in Fig.~\ref{fig:comparisonplot} based on
full-dimensional calculations; at first sight, the single dominant progression
suggests that only one degree of freedom may be responsible for the major
features of the spectra. Exact quantum calculations for the one-dimensional
model whose details are available in the Supporting Information will help shed
light on this apparent simplicity.

\begin{figure}
[H]\centering\includegraphics[scale=1]{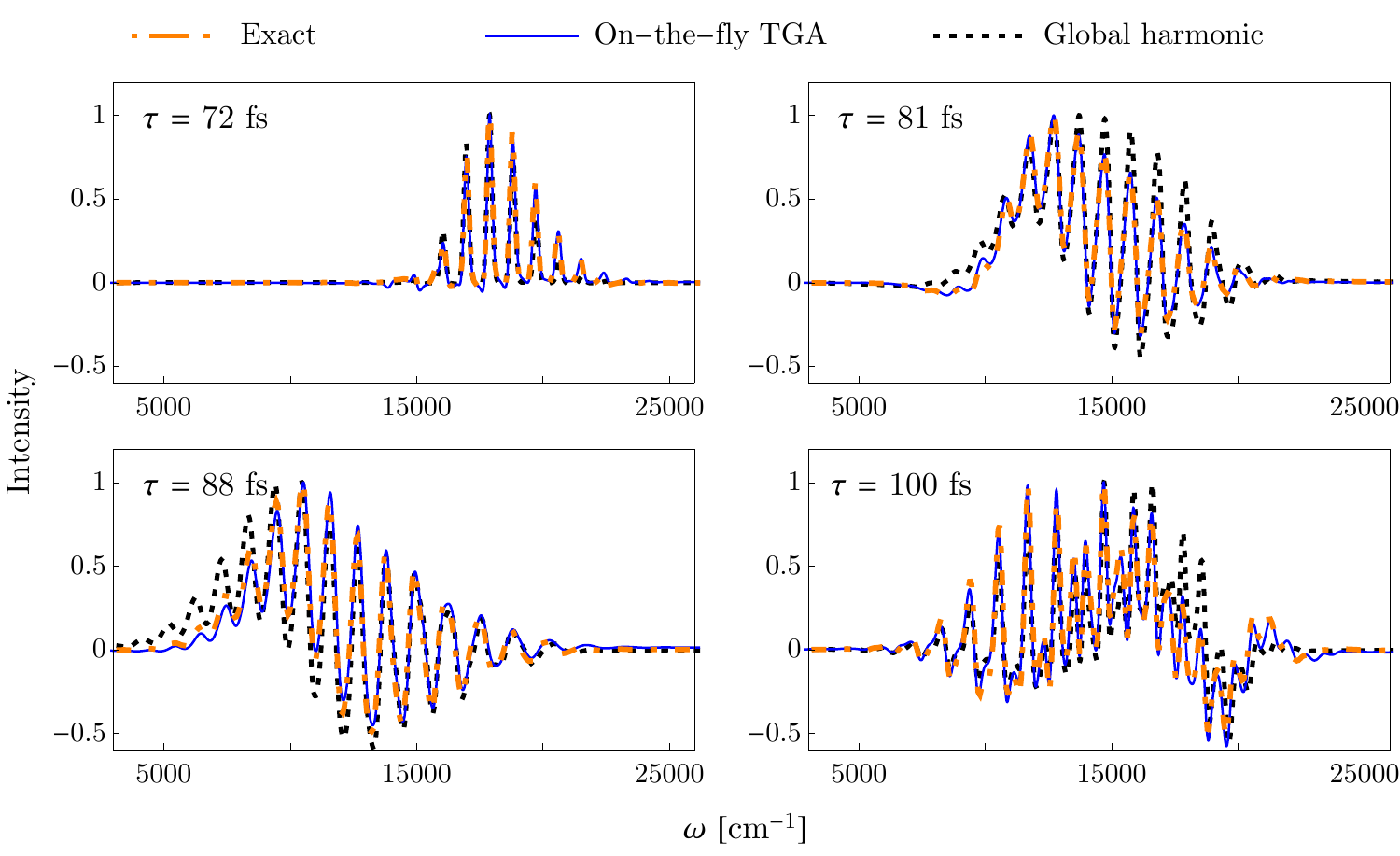}
\caption{\label{fig:ProbesExact}Time-resolved spectra of the one-dimensional model of phenyl radical
evaluated using the exact quantum dynamics, on-the-fly thawed Gaussian approximation (TGA), and global
harmonic approach. Probe delay times are taken as to follow the wavepacket motion during one period
of $\approx36$ fs between the second and third recurrences. }
\end{figure}

Figure~\ref{fig:ProbesExact} compares the time-resolved spectra of the
one-dimensional model evaluated with the exact quantum dynamics, on-the-fly
thawed Gaussian approximation, and adiabatic global harmonic approach. The
spectra are presented at the probe delay times which follow the wavepacket
during one period of oscillation between the second and third recurrences,
i.e., after a rather long anharmonic dynamics on the excited-state surface.
The thawed Gaussian approximation performs very well at this level of
anharmonicity and for the given time scales, almost perfectly reproducing the
exact spectra at all probe delay times, even at times when the wavepacket is
far from the Franck--Condon region. Such accuracy requires that the
excited-state wavepacket be well described by a Gaussian---indeed, we find
only minor difference between the exact and thawed Gaussian wavepackets (see
Fig.~3 of the Supporting Information). In contrast to the thawed Gaussian
approximation, the global harmonic model fails to recover the exact result.
Although the global harmonic model yields only slightly incorrect peak
intensities in the steady-state absorption spectrum (see Fig.~4 of the
Supporting Information), these differences are enhanced in the time-resolved
spectra in Fig.~\ref{fig:ProbesExact}, confirming that pump-probe spectroscopy
is a more challenging task for approximate methods.

The spectra of Fig.~\ref{fig:ProbesExact} exhibit a similar pattern as the
full-dimensional calculations in Fig.~\ref{fig:comparisonplot}, namely the
periodic spreading and narrowing of the spectra, as well as the time-dependent
displacement of the spectra following the energy gap between the two surfaces.
Nevertheless, the spectral features differ substantially from those observed
in the full-dimensional calculation (Fig.~\ref{fig:comparisonplot}). First,
the one-dimensional model cannot reproduce the loss of vibrational resolution
at the delay times when the wavepacket is far from the Franck--Condon region.
Second, the broad baseline of the spectra computed using the full-dimensional
approach is missing. Third, the double-headed progression, found in the
steady-state absorption spectrum of the phenyl radical\cite{Radziszewski:1999}
and observed at some delay times (e.g. at $\tau=72$ fs, see
Fig.~\ref{fig:comparisonplot}) does not appear in the one-dimensional results.

Finally, to probe the limits of the thawed Gaussian approximation, we
constructed a more challenging example, where the parameters of the Morse
potential were modified in order to enhance the anharmonicity. Whereas the
resulting thawed-Gaussian spectra (see Fig.~5 of the Supporting Information)
are still qualitatively correct, they do differ from the exact spectra due to
the increased anharmonicity. The thawed Gaussian approximation provides good
estimates of the center and width of the wavepacket evolving in the excited
electronic state (see Fig.~6 of the Supporting Information); however, even the
minor differences between the exact and thawed Gaussian wavepackets affect the
propagation on the ground-state surface. As expected, the differences in the
propagated wavepackets arise due to the inability of the thawed Gaussian
approximation to describe the asymmetric deformations of the wavepacket
evolved in an anharmonic potential. The time-resolved spectrum evaluated using
the adiabatic harmonic model fails completely, even though its linear
absorption spectrum agrees with the exact result at least qualitatively (see
Fig.~7 of the Supporting Information). In summary, these results confirm that
the thawed Gaussian approximation starts to break down much later than does
the global harmonic approach; in particular, the thawed Gaussian approximation
can remain valid in realistic anharmonic systems, in which the global harmonic
approach already fails even qualitatively.

\section{Conclusion}

To conclude, we presented an efficient on-the-fly \emph{ab initio} method for
computing vibrational structure of time-resolved electronic spectra. The
method, based on the thawed Gaussian approximation, does not require expensive
construction of potential energy surfaces, and is, therefore, applicable to
high-dimensional problems. In addition, the method avoids making any
assumptions about the \emph{global} shape of the potential energy surfaces and
uses only local dynamical information.Yet, the inherent and severe
approximation, assuming a Gaussian form of the wavepacket, is expected to fail
in more anharmonic potentials. Therefore, we complemented our \emph{ab initio}
calculations of the time-resolved spectra of the phenyl radical by comparing
the thawed Gaussian approximation with the exact solution of a one-dimensional
model based on the most-displaced vibrational mode of the phenyl radical. The
one-dimensional results support the \emph{ab initio }calculations by
confirming that (i) the thawed Gaussian approximation remains valid in a
one-dimensional system with comparable anharmonicity and (ii) that the
on-the-fly \emph{ab initio} spectra can only be reproduced by including
multiple degrees of freedom and couplings between them.

To evaluate the effect of anharmonicity on the pump-probe spectra, we compared
the on-the-fly approach with two approximate methods based on the global
harmonic approximation. The results confirm that the anharmonicity of the
excited-state potential has different effects on the spectrum, depending on
the probe delay time---at delays for which the wavepacket is near the
Franck--Condon region, the anharmonicity influences the relative intensities
of well-resolved vibronic peaks, whereas at delays for which the wavepacket is
located far from the initial geometry, the vibrational resolution is lost and
the anharmonicity only modifies the mean and width of the spectrum.
Interestingly, the two approximate approaches---one assuming quadratic
potential for both surfaces, the other using the global harmonic approximation
only for the ground electronic state---break down in different regimes,
depending on the position of the nuclear wavepacket at the probe delay time.
The combined on-the-fly/global harmonic method presented here serves as a
measure of the anharmonicity effects in both potential energy surfaces; while
the on-the-fly and global harmonic approaches give qualitatively similar
results, indicating only minor anharmonicity of the potential energy surfaces,
the combined approach breaks down completely for time delays when the
wavepacket is far from the Franck--Condon point, thus confirming that the
global harmonic approximation works only because of cancellation of errors.
These insights, provided by the full-dimensional calculations based on the
on-the-fly \emph{ab initio} thawed Gaussian approximation, can help build more
computationally efficient approaches to evaluating time-resolved spectra with
rather high accuracy and minimum human input.

The method presented here is unable to describe arbitrary pulse shapes and
durations---this would require the computation of the full third-order
response functions and a subsequent integration over the pump and probe
pulses.\cite{Pollard_Mathies:1992} Such scheme could be combined with the
thawed Gaussian propagation of the wavepacket but would require much smaller
spacings between the probe delay times in order to perform the integration.
Another limitation of the presented approach is the wavefunction formalism,
which is not suited for accounting for finite-temperature or environment
effects---this could be augmented by recasting the thawed Gaussian
approximation into the density matrix formalism, as recently shown for the
harmonic oscillator by Reddy and Prasad.\cite{Reddy_Prasad:2016} Both
extensions are envisioned and will allow for more direct comparison with experiment.

\section*{Supplementary material}

See supplementary material for information on the most displaced modes of the
phenyl radical, details of the one-dimensional \emph{ab initio} model and
associated quantum calculations, figures of the excited-state wavepacket,
absorption and time-resolved stimulated spectra of the one-dimensional
\emph{ab initio} model as well as of a model with increased anharmonicity.

\begin{acknowledgments}
The authors acknowledge the financial support from the Swiss National Science
Foundation through the NCCR MUST (Molecular Ultrafast Science and Technology)
Network, from the European Research Council (ERC) under the European Union's
Horizon 2020 research and innovation programme (grant agreement No. 683069 --
MOLEQULE), and from the COST action MOLIM (Molecules in Motion).
\end{acknowledgments}

\appendix

\section{\label{sec_app:C_mu_tau}Expanded form of the dipole time
autocorrelation function of Eq.~(\ref{eq:C_mu_tau})}

Here we show the fully expanded form of Eq.~(\ref{eq:C_mu_tau}) that is used
in subsequent derivation of non-vanishing terms in Appendix
\ref{sec_app:sigma_PP_CA_final}. The expansion of the commutator of
Eq.~(\ref{eq:rho_pu}) gives three terms:
\begin{equation}
\bm{\hat{\rho}}^{\text{pu}}=2\pi^{2}\hbar^{-2}(2\bm{\hat{\mu}}_{\text{RC}%
}^{\text{pu}}\bm{\hat{\rho}}\bm{\hat{\mu}}_{\text{RC}}^{\text{pu}%
}-\bm{\hat{\mu}}_{\text{RC}}^{\text{pu}}\bm{\hat{\mu}}_{\text{RC}}^{\text{pu}%
}\bm{\hat{\rho}}-\bm{\hat{\rho}}\bm{\hat{\mu}}_{\text{RC}}^{\text{pu}%
}\bm{\hat{\mu}}_{\text{RC}}^{\text{pu}}). \label{eq:rho_pu_full}%
\end{equation}
Inserting the expression (\ref{eq:rho_pu_full}) into Eq.~(\ref{eq:C_mu_tau}),
expanding the dipole operator [Eq.~(\ref{eq:mu_pr_t})], and using the cyclic
property of the trace yields:
\begin{align}
C_{\mu,\tau}^{(3)}(t^{\prime}) = 2 \pi^{2} \hbar^{-2} \biggl\{2  &
\text{Tr}\left[  \bm{\hat{\rho}}\bm{\hat{\mu}}_{\text{RC}}^{\text{pu}%
}e^{i\mathbf{\hat{H}}\tau/\hbar}\bm{\hat{\mu}}^{\text{pr}}e^{i\mathbf{\hat{H}%
}t^{\prime}/\hbar}\bm{\hat{\mu}}^{\text{pr}}e^{-i\mathbf{\hat{H}}(t^{\prime
}+\tau)/\hbar}\bm{\hat{\mu}}_{\text{RC}}^{\text{pu}}\right] \nonumber\\
-  &  \text{Tr}\left[  \bm{\hat{\rho}}\bm{\hat{\mu}}_{\text{RC}}^{\text{pu}%
}\bm{\hat{\mu}}_{\text{RC}}^{\text{pu}}e^{i\mathbf{\hat{H}}\tau/\hbar
}\bm{\hat{\mu}}^{\text{pr}}e^{i\mathbf{\hat{H}}t^{\prime}/\hbar}%
\bm{\hat{\mu}}^{\text{pr}}e^{-i\mathbf{\hat{H}}(t^{\prime}+\tau)/\hbar}\right]
\label{eq:C_mu_tau_full}\\
-  &  \text{Tr}\left[  \bm{\hat{\rho}}e^{i\mathbf{\hat{H}}\tau/\hbar
}\bm{\hat{\mu}}^{\text{pr}}e^{i\mathbf{\hat{H}}t^{\prime}/\hbar}%
\bm{\hat{\mu}}^{\text{pr}}e^{-i\mathbf{\hat{H}}(t^{\prime}+\tau)/\hbar
}\bm{\hat{\mu}}_{\text{RC}}^{\text{pu}}\bm{\hat{\mu}}_{\text{RC}}^{\text{pu}%
}\right]  \biggr\}\,.\nonumber
\end{align}

\section{\label{sec_app:sigma_PP_CA_final}Derivation of the differential
absorption cross-section of Eq.~(\ref{eq:sigma_PP_CA_final})}

To show explicitly which terms in Eq.~(\ref{eq:sigma_PP}) vanish and which
contribute to the spectrum within the additional approximations imposed, it is
convenient, for each electronic state $j$, to separate the
coordinate-independent, electronic Hamiltonian component, given by the
vertical transition energy $\hbar\omega_{j0}$, from the coordinate-dependent,
nuclear Hamiltonian $\hat{h}_{j}$:
\begin{equation}
\hat{H}_{j}=\hbar\omega_{j0}+\hat{h}_{j}\,,\,j=0,1,2\,. \label{eq:h_j}%
\end{equation}
In the remainder, we also use general $\hbar\omega_{ij}$ to denote the energy
gap between electronic states $i$ and $j$ at the initial geometry. This allows
us to group and analyze the highly oscillatory terms separately from the
slowly-varying nuclear contribution to the spectrum.

The first term of the dipole time autocorrelation function
[Eq.~(\ref{eq:C_mu_tau_full})] is expanded as
\begin{align}
&  \text{Tr}[\bm{\hat{\rho}}\bm{\hat{\mu}}_{\text{RC}}^{\text{pu}%
}e^{i\mathbf{\hat{H}}\tau/\hbar}\bm{\hat{\mu}}^{\text{pr}}e^{i\mathbf{\hat{H}%
}t^{\prime}/\hbar}\bm{\hat{\mu}}^{\text{pr}}e^{-i\mathbf{\hat{H}}(t^{\prime
}+\tau)/\hbar}\bm{\hat{\mu}}_{\text{RC}}^{\text{pu}}]\nonumber\\
&  =|\tilde{E}^{\text{pu}}(\omega_{10})|^{2}\langle\psi_{0,g}|\hat{\mu}%
_{01}^{\text{pu}}e^{i\hat{H}_{1}\tau/\hbar}\hat{\mu}_{10}^{\text{pr}}%
e^{i\hat{H}_{0}t^{\prime}/\hbar}\hat{\mu}_{01}^{\text{pr}}e^{-i\hat{H}%
_{1}(t^{\prime}+\tau)/\hbar}\hat{\mu}_{10}^{\text{pu}}|\psi_{0,g}%
\rangle\nonumber\\
&  +|\tilde{E}^{\text{pu}}(\omega_{10})|^{2}\langle\psi_{0,g}|\hat{\mu}%
_{01}^{\text{pu}}e^{i\hat{H}_{1}\tau/\hbar}\hat{\mu}_{12}^{\text{pr}}%
e^{i\hat{H}_{2}t^{\prime}/\hbar}\hat{\mu}_{21}^{\text{pr}}e^{-i\hat{H}%
_{1}(t^{\prime}+\tau)/\hbar}\hat{\mu}_{10}^{\text{pu}}|\psi_{0,g}\rangle\\
&  =|\tilde{E}^{\text{pu}}(\omega_{10})|^{2}e^{-i\omega_{10}t^{\prime}}%
\langle\psi_{0,g}|\hat{\mu}_{01}^{\text{pu}}e^{i\hat{h}_{1}\tau/\hbar}\hat
{\mu}_{10}^{\text{pr}}e^{i\hat{h}_{0}t^{\prime}/\hbar}\hat{\mu}_{01}%
^{\text{pr}}e^{-i\hat{h}_{1}(t^{\prime}+\tau)/\hbar}\hat{\mu}_{10}^{\text{pu}%
}|\psi_{0,g}\rangle\nonumber\\
&  +|\tilde{E}^{\text{pu}}(\omega_{10})|^{2}e^{i\omega_{21}t^{\prime}}%
\langle\psi_{0,g}|\hat{\mu}_{01}^{\text{pu}}e^{i\hat{h}_{1}\tau/\hbar}\hat
{\mu}_{12}^{\text{pr}}e^{i\hat{h}_{2}t^{\prime}/\hbar}\hat{\mu}_{21}%
^{\text{pr}}e^{-i\hat{h}_{1}(t^{\prime}+\tau)/\hbar}\hat{\mu}_{10}^{\text{pu}%
}|\psi_{0,g}\rangle. \label{eq:term1_last}%
\end{align}
In the first step, we have applied the three-state system assumption,
zero-temperature limit (constraint on $\bm{\hat{\rho}}$), and
Born--Oppenheimer approximation (there is no population transfer during the
field-free evolution), while in the second step, we have used the separation
Eq.~(\ref{eq:h_j}). The first term of (\ref{eq:term1_last}) gives the
time-resolved stimulated emission---its Fourier transform is centered near
$\omega_{10}$ and modulated by the dynamics on the ground and first excited
states. In contrast, the second term of Eq. (\ref{eq:term1_last}) vanishes
within the rotating-wave approximation---its Fourier transform would lead to a
spectrum at negative frequencies around $-\omega_{21}$. We recall now that the
full expression for the spectrum [Eq.~(\ref{eq:sigma_PP})] involves also the
complex conjugate of the correlation function, for which the vanishing term is
the complex conjugate of the first term of Eq.~(\ref{eq:term1_last}), whereas
the complex conjugate of the second term of Eq.~(\ref{eq:term1_last}) is
nonzero and gives rise to the time-resolved excited-state absorption spectrum.

Similarly, the second term in Eq.~(\ref{eq:C_mu_tau_full}) is
\begin{align}
&  \text{Tr}[\bm{\hat{\rho}}\bm{\hat{\mu}}_{\text{RC}}^{\text{pu}%
}\bm{\hat{\mu}}_{\text{RC}}^{\text{pu}}e^{i\mathbf{\hat{H}}\tau/\hbar
}\bm{\hat{\mu}}^{\text{pr}}e^{i\mathbf{\hat{H}}t^{\prime}/\hbar}%
\bm{\hat{\mu}}^{\text{pr}}e^{-i\mathbf{\hat{H}}(t^{\prime}+\tau)/\hbar
}]\nonumber\\
&  =|\tilde{E}^{\text{pu}}(\omega_{10})|^{2}\langle\psi_{0,g}|\hat{\mu}%
_{01}^{\text{pu}}\hat{\mu}_{10}^{\text{pu}}e^{i\hat{H}_{0}\tau/\hbar}\hat{\mu
}_{01}^{\text{pr}}e^{i\hat{H}_{1}t^{\prime}/\hbar}\hat{\mu}_{10}^{\text{pr}%
}e^{-i\hat{H}_{0}(t^{\prime}+\tau)/\hbar}|\psi_{0,g}\rangle\nonumber\\
&  +|\tilde{E}^{\text{pu}}(\omega_{10})|^{2}\langle\psi_{0,g}|\hat{\mu}%
_{01}^{\text{pu}}\hat{\mu}_{12}^{\text{pu}}e^{i\hat{H}_{2}\tau/\hbar}\hat{\mu
}_{21}^{\text{pr}}e^{i\hat{H}_{1}t^{\prime}/\hbar}\hat{\mu}_{10}^{\text{pr}%
}e^{-i\hat{H}_{0}(t^{\prime}+\tau)/\hbar}|\psi_{0,g}\rangle\\
&  =|\tilde{E}^{\text{pu}}(\omega_{10})|^{2}e^{i\omega_{10}t^{\prime}}%
\langle\psi_{0,g}|\hat{\mu}_{01}^{\text{pu}}\hat{\mu}_{10}^{\text{pu}}%
e^{i\hat{h}_{0}\tau/\hbar}\hat{\mu}_{01}^{\text{pr}}e^{i\hat{h}_{1}t^{\prime
}/\hbar}\hat{\mu}_{10}^{\text{pr}}e^{-i\hat{h}_{0}(t^{\prime}+\tau)/\hbar
}|\psi_{0,g}\rangle\nonumber\\
&  +|\tilde{E}^{\text{pu}}(\omega_{10})|^{2}e^{i\omega_{20}\tau}%
e^{i\omega_{10}t^{\prime}}\langle\psi_{0,g}|\hat{\mu}_{01}^{\text{pu}}\hat
{\mu}_{12}^{\text{pu}}e^{i\hat{h}_{2}\tau/\hbar}\hat{\mu}_{21}^{\text{pr}%
}e^{i\hat{h}_{1}t^{\prime}/\hbar}\hat{\mu}_{10}^{\text{pr}}e^{-i\hat{h}%
_{0}(t^{\prime}+\tau)/\hbar}|\psi_{0,g}\rangle\,. \label{eq:term2_last}%
\end{align}
The Fourier transforms of both terms of Eq.~(\ref{eq:term2_last}) are zero
within the rotating-wave approximation. However, this does not hold for their
complex conjugates---the complex conjugate of the first term of
Eq.~(\ref{eq:term2_last}) gives rise to the ground-state bleach signal---it is
centered near $\omega_{10}$, but unlike the time-resolved stimulated emission,
does not involve wavepacket propagation on the excited electronic state during
the delay time $\tau$. Regarding the complex conjugate of the second term in
Eq.~(\ref{eq:term2_last}), we recall that the pulses are short on the nuclear
time scale, which allows the use of the ultrashort pulse approximation, but
not on the electronic scale---the averaging of the probe spectra over the
short probe pulse duration leads to the cancellation of the terms with a
highly oscillatory phase $\exp(\pm i \omega_{20} \tau)$.

For the last term of Eq.~(\ref{eq:C_mu_tau_full}) we obtain
\begin{align}
&  \text{Tr}[\bm{\hat{\rho}}e^{i\mathbf{\hat{H}}\tau/\hbar}%
\bm{\hat{\mu}}^{\text{pr}}e^{i\mathbf{\hat{H}}t^{\prime}/\hbar}%
\bm{\hat{\mu}}^{\text{pr}}e^{-i\mathbf{\hat{H}}(t^{\prime}+\tau)/\hbar
}\bm{\hat{\mu}}_{\text{RC}}^{\text{pu}}\bm{\hat{\mu}}_{\text{RC}}^{\text{pu}%
}]\nonumber\\
&  =|\tilde{E}^{\text{pu}}(\omega_{10})|^{2}\langle\psi_{0,g}|e^{i\hat{H}%
_{0}\tau/\hbar}\hat{\mu}_{01}^{\text{pr}}e^{i\hat{H}_{1}t^{\prime}/\hbar}%
\hat{\mu}_{10}^{\text{pr}}e^{-i\hat{H}_{0}(t^{\prime}+\tau)/\hbar}\hat{\mu
}_{01}^{\text{pu}}\hat{\mu}_{10}^{\text{pu}}|\psi_{0,g}\rangle\nonumber\\
&  +|\tilde{E}^{\text{pu}}(\omega_{10})|^{2}\langle\psi_{0,g}|e^{i\hat{H}%
_{0}\tau/\hbar}\hat{\mu}_{01}^{\text{pr}}e^{i\hat{H}_{1}t^{\prime}/\hbar}%
\hat{\mu}_{12}^{\text{pr}}e^{-i\hat{H}_{2}(t^{\prime}+\tau)/\hbar}\hat{\mu
}_{21}^{\text{pu}}\hat{\mu}_{10}^{\text{pu}}|\psi_{0,g}\rangle\\
&  =|\tilde{E}^{\text{pu}}(\omega_{10})|^{2}e^{i\omega_{10}t^{\prime}}%
\langle\psi_{0,g}|e^{i\hat{h}_{0}\tau/\hbar}\hat{\mu}_{01}^{\text{pr}}%
e^{i\hat{h}_{1}t^{\prime}/\hbar}\hat{\mu}_{10}^{\text{pr}}e^{-i\hat{h}%
_{0}(t^{\prime}+\tau)/\hbar}\hat{\mu}_{01}^{\text{pu}}\hat{\mu}_{10}%
^{\text{pu}}|\psi_{0,g}\rangle\nonumber\\
&  +|\tilde{E}^{\text{pu}}(\omega_{10})|^{2}e^{-i\omega_{21}t^{\prime}%
}e^{-i\omega_{20}\tau}\langle\psi_{0,g}|e^{i\hat{h}_{0}\tau/\hbar}\hat{\mu
}_{01}^{\text{pr}}e^{i\hat{h}_{1}t^{\prime}/\hbar}\hat{\mu}_{12}^{\text{pr}%
}e^{-i\hat{h}_{2}(t^{\prime}+\tau)/\hbar}\hat{\mu}_{21}^{\text{pu}}\hat{\mu
}_{10}^{\text{pu}}|\psi_{0,g}\rangle\,. \label{eq:term3_last}%
\end{align}
Again, the first term of Eq.~(\ref{eq:term3_last}) vanishes due to the
rotating-wave approximation, while its complex conjugate does not vanish and
corresponds to the ground-state bleach. The second term of
Eq.~(\ref{eq:term3_last}) vanishes due to the highly oscillatory term $\exp(-
i\omega_{20}\tau)$ and its complex conjugate vanishes both due to the
rotating-wave approximation and the oscillatory term $\exp(i\omega_{20}\tau)$.

We see that the first term of Eq.~(\ref{eq:C_mu_tau_full}) gives the
time-resolved stimulated emission and excited-state absorption contributions,
while the other two terms contribute to the ground-state bleach through two
distinct expressions. If one further assumes that the electronic transition
dipole moment elements are independent of the nuclear coordinates, i.e., the
Condon approximation, the two contributions to the ground-state bleach are
equal and correspond to the scaled linear absorption spectrum. Finally,
summing all non-zero terms and reintroducing the nuclear Hamiltonians from
Eq.~(\ref{eq:h_j}) gives the expressions (\ref{eq:sigma_PP_CA_final}%
)--(\ref{eq:GSB}).

\section{\label{sec_app:orient_av}Orientational averaging of pump-probe
spectra}

Here we derive the constant scaling factors for the time-resolved spectrum of
Eq.~(\ref{eq:sigma_PP_CA_final}) arising from averaging over all possible
orientations of molecules in an isotropic sample. We first consider a rank-4
three-dimensional tensor $\overleftrightarrow{\sigma}(\omega)$ related to the
spectrum by
\begin{equation}
\sigma(\omega)=\epsilon_{I}^{\text{pu}}\epsilon_{J}^{\text{pr}}\epsilon
_{K}^{\text{pr}}\epsilon_{L}^{\text{pu}}\overleftrightarrow{\sigma}%
_{IJKL}(\omega), \label{eq:sigma_tensor}%
\end{equation}
where Einstein's summation convention was used (i.e., a sum over
$I,J,K,L=1,2,3$ is implied). For fixed polarization vectors of the pump
($\vec{\epsilon}^{\,\text{pu}}$) and probe ($\vec{\epsilon}^{\,\text{pr}}$)
pulses in the laboratory frame, one can easily evaluate the spectrum from
Eq.~(\ref{eq:sigma_tensor}). However, the spectrum tensor is a molecular
property, and as such, is more naturally given in the molecular frame as
$\overleftrightarrow{\sigma}_{ijkl}(\omega)$, where lowercase indices are used
to distinguish the molecular from laboratory frame. Tensors in the two frames
are related by%
\begin{equation}
\overleftrightarrow{\sigma}_{IJKL}(\omega)=\lambda_{Ii}\lambda_{Jj}%
\lambda_{Kk}\lambda_{Ll}\overleftrightarrow{\sigma}_{ijkl}(\omega),
\label{eq:sigma_mol_to_lab}%
\end{equation}
where the coefficients $\lambda_{Ii},\lambda_{Jj},...$ represent the cosines
of angles between the laboratory and molecular axes indicated in the
subscript. The orientational average of $\overleftrightarrow{\sigma}%
_{IJKL}(\omega)$ is obtained by averaging the coefficients $\lambda
_{Ii}\lambda_{Jj}\lambda_{Kk}\lambda_{Ll}$. Following the procedure of Craig
and Thirunamachandran,\cite{Craig_Thirunamachandran:1984} but using a slightly
more general expression from Ref.~\onlinecite{Hamm_Zanni:2011}, one obtains
the average
\begin{equation}
\overline{\lambda_{Ii}\lambda_{Jj}\lambda_{Kk}\lambda_{Ll}}=\frac{1}{30}%
\begin{pmatrix}
\lambda_{IJ}\lambda_{KL}\\
\lambda_{IK}\lambda_{JL}\\
\lambda_{IL}\lambda_{JK}%
\end{pmatrix}
^{T}\cdot%
\begin{pmatrix}
4 & -1 & -1\\
-1 & 4 & -1\\
-1 & -1 & 4
\end{pmatrix}
\cdot%
\begin{pmatrix}
\lambda_{ij}\lambda_{kl}\\
\lambda_{ik}\lambda_{jl}\\
\lambda_{il}\lambda_{jk}%
\end{pmatrix}
. \label{eq:orient_av_Hamm_Zanni}%
\end{equation}

Now we fix the laboratory frame so that the pump pulse is polarized along the
$Z$-axis, and let the probe pulse be polarized along an axis, denoted $\alpha
$, tilted by an angle $\alpha$ with respect to the $Z$-axis. Then the
sum~(\ref{eq:sigma_tensor}) reduces to a single term
\begin{equation}
\sigma(\omega)=\overleftrightarrow{\sigma}_{Z\alpha\alpha Z}(\omega).
\end{equation}
The molecular frame is set so that the $z$-axis is parallel to the transition
dipole moment vector. Within the Condon approximation, only one component of
the spectrum tensor in the molecular frame is nonzero and the
Eq.~(\ref{eq:sigma_mol_to_lab}) simplifies to
\begin{equation}
\overleftrightarrow{\sigma}_{Z\alpha\alpha Z,\,\text{CA}}(\omega)=\lambda
_{Zz}\lambda_{\alpha z}\lambda_{\alpha z}\lambda_{Zz}%
\overleftrightarrow{\sigma}_{zzzz,\,\text{CA}}(\omega).
\label{eq:CA_orient_av}%
\end{equation}
In general, the transition dipole moment is a complicated function of the
geometry and can change its orientation during the probe delay time, which
gives rise to other nonzero spectrum tensor elements. The reorientation of the
transition dipole moment can also occur due to the rotation of the molecule,
which is, however, neglected here because of the ultrashort time scales
involved. Applying Eq.~(\ref{eq:orient_av_Hamm_Zanni}) gives:
\begin{align}
\overline{\lambda_{Zz}\lambda_{\alpha z}\lambda_{\alpha z}\lambda_{Zz}}  &
=\frac{1}{30}%
\begin{pmatrix}
\cos^{2}\alpha\\
\cos^{2}\alpha\\
1
\end{pmatrix}
^{T}\cdot%
\begin{pmatrix}
4 & -1 & -1\\
-1 & 4 & -1\\
-1 & -1 & 4
\end{pmatrix}
\cdot%
\begin{pmatrix}
1\\
1\\
1
\end{pmatrix}
\\
&  =\frac{1}{30}(4\cos^{2}\alpha+2)\\
&  =\frac{1}{5}\cos^{2}\alpha+\frac{1}{15}\sin^{2}\alpha.
\end{align}
Thus, the orientationally averaged time-resolved spectrum is 5 to 15 times
weaker than the spectrum that would be obtained if the polarizations of both
the pump and probe electric fields were aligned with the transition dipole
moment of the molecule---the exact ratio depends on the angle $\alpha$ between
the pump and probe polarization vectors. Within the Condon approximation, the
orientation average is needed only for absolute cross sections; if only
relative cross sections are required, the orientational averaging is
unnecessary because the effect of averaging is equivalent to a scaling of the
spectra by a constant factor.

\section{\label{sec_app:position_width}Derivation of the expressions for the
means and widths of the time-resolved stimulated emission spectra}

Let $a$, $\kappa$, $\kappa_{1}$, and $\kappa_{2}$ be $D\times D$ real
symmetric matrices and let $a$ be positive definite; in our setting, $D$
denotes the number of degrees of freedom. In this Appendix, we will use the
following analytical expressions for three Gaussian
integrals:\cite{Brookes:2011,Petersen_Pedersen:2012,Sulc_Vanicek:2013}
\begin{align}
I_{0}(a)  &  :=\int d^{D}qe^{-q^{T}\cdot a\cdot q}=\left(  \frac{\pi^{D}}{\det
a}\right)  ^{\frac{1}{2}},\label{eq:I_0}\\
I_{2}(a,\kappa)  &  :=\int d^{D}q(q^{T}\cdot\kappa\cdot q)e^{-q^{T}\cdot
a\cdot q}\nonumber\\
&  =\frac{1}{2}I_{0}(a)\text{Tr}(\kappa\cdot a^{-1}),\label{eq:I_2}\\
I_{4}(a,\kappa_{1},\kappa_{2})  &  :=\int d^{D}q(q^{T}\cdot\kappa_{1}\cdot
q)(q^{T}\cdot\kappa_{2}\cdot q)e^{-q^{T}\cdot a\cdot q}\nonumber\\
&  =\frac{1}{2}I_{0}(a)\left[  \frac{1}{2}\text{Tr}(\kappa_{1}\cdot
a^{-1})\text{Tr}(\kappa_{2}\cdot a^{-1})+\text{Tr}(\kappa_{1}\cdot a^{-1}%
\cdot\kappa_{2}\cdot a^{-1})\right]  . \label{eq:I_4}%
\end{align}

Within the local harmonic approximation, the difference between the effective
time-dependent potentials of the two surfaces is
\begin{equation}
\Delta V_{\text{eff}}(q,\tau)=\Delta V_{\tau}+(\Delta V^{\prime}_{\tau}%
)^{T}\cdot(q-q_{\tau})+\frac{1}{2}(q-q_{\tau})^{T}\cdot\Delta V^{\prime\prime
}_{\tau}\cdot(q-q_{\tau}), \label{eq:deltaV}%
\end{equation}
where we used definitions~(\ref{eq:delta_V_tau})--(\ref{eq:delta_H_tau}) of
$\Delta V_{\tau}$, $\Delta V^{\prime}_{\tau}$, and $\Delta V^{\prime\prime
}_{\tau}$. To evaluate the mean of the normalized time-resolved stimulated
emission spectrum (\ref{eq:sigma_normalized}), we start from
Eq.~(\ref{eq:mean_wp}) and insert the expression for the potential difference
(\ref{eq:deltaV}):
\begin{equation}
\langle\omega\rangle_{\tau}^{\text{LHA}}=\hbar^{-1}\langle\psi_{\tau}%
|\Delta\hat{V}_{\text{eff}}(\tau)|\psi_{\tau}\rangle.
\end{equation}
Within the thawed Gaussian approximation the probability density $|\psi_{\tau
}|^{2}$ is even, so the terms containing odd powers of $(q-q_{\tau})$ vanish
in the integral over all space and the expression reduces to
\begin{align}
\hbar\langle\omega\rangle_{\tau}^{\text{LHA}}  &  =\Delta V_{\tau}+\frac{1}%
{2}\int d^{D}q|\psi_{\tau}|^{2}(q-q_{\tau})^{T}\cdot\Delta V^{\prime\prime
}_{\tau}\cdot(q-q_{\tau})\\
&  =\Delta V_{\tau}+\frac{1}{2}N_{0}^{2}\int d^{D}qe^{-2(q-q_{\tau})^{T}%
\cdot\text{Re}A_{\tau}\cdot(q-q_{\tau})}(q-q_{\tau})^{T}\cdot\Delta
V^{\prime\prime}_{\tau}\cdot(q-q_{\tau}).
\end{align}
We now trivially transform the integration over $q$ to integration over
$(q-q_{\tau})$ and use Eq.~(\ref{eq:I_2}) to obtain Eq.~(\ref{eq:mean_LHA}).

As for the spectral widths, we first have to evaluate the second moment of the
normalized spectral lineshape, which is given by\cite{Pollard_Mathies:1990a}
\begin{equation}
\langle\omega^{2}\rangle_{\tau}=\hbar^{-2}\langle\psi_{\tau}|\Delta\hat{V}%
^{2}|\psi_{\tau}\rangle. \label{eq:second_moment}%
\end{equation}
Inserting the effective local harmonic potential into
Eq.~(\ref{eq:second_moment}) and dropping terms with odd powers of
$(q-q_{\tau})$ yields
\begin{align}
\hbar^{2}\langle\omega^{2}\rangle_{\tau}^{\text{LHA}}=\Delta V_{\tau}^{2}  &
+\int dq|\psi_{\tau}|^{2}\{[(\Delta V^{\prime}_{\tau})^{T}\cdot(q-q_{\tau
})]^{2}\nonumber\\
&  +(q-q_{\tau})^{T}\cdot\Delta V^{\prime\prime}_{\tau}\cdot(q-q_{\tau})
\Delta V_{\tau}\nonumber\\
&  +\frac{1}{4}[(q-q_{\tau})^{T}\cdot\Delta V^{\prime\prime}_{\tau}%
\cdot(q-q_{\tau})]^{2}\}.
\end{align}
Evaluating the integral using Eqs.~(\ref{eq:I_2}) and (\ref{eq:I_4}) gives
\begin{align}
\hbar^{2}\langle\omega^{2}\rangle_{\tau}^{\text{LHA}}  &  =\Delta V_{\tau}%
^{2}+ (\Delta V^{\prime}_{\tau})^{T}\cdot\Sigma_{\tau}^{2}\cdot\Delta
V^{\prime}_{\tau}\nonumber\\
&  + \Delta V_{\tau}\text{Tr}(C)+\frac{1}{4}\left(  \text{Tr}\,C\right)
^{2}+\frac{1}{2}\text{Tr}(C^{2})\\
&  =\hbar^{2}(\langle\omega\rangle_{\tau}^{\text{LHA}})^{2}+ (\Delta
V^{\prime}_{\tau})^{T}\cdot\Sigma_{\tau}^{2}\cdot\Delta V^{\prime}_{\tau
}+\frac{1}{2}\text{Tr}(C^{2}) \label{eq:second_moment_LHA}%
\end{align}
where $\Sigma_{\tau}^{2}:=\left(  1/4\right)  ($Re$A_{\tau})^{-1}$, $C:=\Delta
V^{\prime\prime}_{\tau}\cdot\Sigma_{\tau}^{2}$, and where we employed
Eq.~(\ref{eq:mean_LHA}). Equation~(\ref{eq:var_LHA}) then follows directly
from Eq.~(\ref{eq:second_moment_LHA}).

\section{\label{sec_app:position_width_broadening}Modification of the
Eqs.~(\ref{eq:mean_wp}) and (\ref{eq:var_wp}) for the spectral width when a
broadening function is applied to the spectrum}

Here we derive expression (\ref{eq:var_alpha}) for the spectral width, used in
our calculations instead of Eqs.~(\ref{eq:var_LHA}) and (\ref{eq:var_approx})
due to the presence of a broadening function. We follow closely the derivation
of Pollard et al.,\cite{Pollard_Mathies:1990a} with the exception that our
$\sigma_{0}$ is already normalized so that its zeroth moment is $1$. The final
expressions for the first and second moments without broadening are equal to
those from Ref.~\onlinecite{Pollard_Mathies:1990a} up to constant factors
which vanish after dividing by the zeroth moment.

The $n$th moment of the normalized spectral lineshape $\sigma_{0}(\omega
,\tau)$ at a probe delay time $\tau$ is given by the derivative of the
corresponding correlation function evaluated at time zero:
\begin{equation}
\langle\omega^{n}\rangle=\text{Re}[i^{n}\frac{d^{n}}{dt^{n}}C(t^{\prime}%
,\tau)|_{t^{\prime}=0}].
\end{equation}
We now take our correlation function to be the stimulated emission correlation
function~(\ref{eq:SE_fidelity}) multiplied by a decay function $f(t^{\prime})$
responsible for the phenomenological broadening of the spectra. Then the
expression for the $n$th moment of the normalized spectral lineshape reads
\begin{equation}
\langle\omega^{n}\rangle=\text{Re}\left[  \sum_{j=0}^{n}i^{n}\binom{n}{j}%
\frac{d^{n-j}}{dt^{\prime n-j}}C_{\text{SE}}(t^{\prime},\tau)|_{t^{\prime}%
=0}\frac{d^{j}}{dt^{\prime j}}f(t^{\prime})|_{t^{\prime}=0}\right]  .
\end{equation}

Before continuing, we note that for the commonly used exponential decay, which
results in the Lorentzian broadening of the spectral lines, the standard
deviation of the spectrum is not defined. We, therefore, used the Gaussian
decay function $f(t^{\prime})=\exp(-\alpha t^{\prime2})$. Since $f(0)=1$ and
$f^{\prime}(0)=0$, the first moment is not affected by this decay function.
However, the second moment changes to
\begin{align}
\langle\omega^{2}\rangle &  =\text{Re}\left[  -\sum_{j=0}^{2}\binom{2}{j}%
\frac{d^{2-j}}{dt^{\prime2-j}}C_{\text{SE}}(t^{\prime},\tau)|_{t^{\prime}%
=0}\frac{d^{j}}{dt^{\prime j}}f(t^{\prime})|_{t^{\prime}=0}\right]
\label{eq:second_moment_broad_1}\\
&  =\text{Re}\left[  -\frac{d^{2}}{dt^{\prime2}}C_{\text{SE}}(t^{\prime}%
,\tau)|_{t^{\prime}=0}-\frac{d^{2}}{dt^{\prime2}}f(t^{\prime})|_{t^{\prime}%
=0}\right] \label{eq:second_moment_broad_2}\\
&  =\frac{1}{\hbar^{2}}\langle\psi_{\tau}|\Delta\hat{V}^{2}|\psi_{\tau}%
\rangle+2\alpha. \label{eq:second_moment_broad}%
\end{align}
To obtain Eq.~(\ref{eq:second_moment_broad}) from
Eq.~(\ref{eq:second_moment_broad_2}), we used the result of Pollard et al. for
the first term, and the second derivative of the Gaussian function,
\begin{equation}
\frac{d^{2}}{dt^{\prime2}}f(t^{\prime})|_{t^{\prime}=0}=(4\alpha^{2}%
t^{\prime2}-2\alpha)f(t^{\prime})|_{t^{\prime}=0}=-2\alpha,
\end{equation}
for the second term.
\bibliographystyle{aipnum4-1}
\bibliography{Pump_probe}

\end{document}


\title{Supporting Information for: On-the-fly \emph{ab initio} semiclassical
evaluation of time-resolved electronic spectra}
\author{Tomislav Begu\v{s}i\'{c}}
\author{Julien Roulet}
\author{Ji\v{r}\'i Van\'i\v{c}ek}
\email{jiri.vanicek@epfl.ch.}
\affiliation{Laboratory of Theoretical Physical Chemistry, Institut des Sciences et
Ing\'enierie Chimiques, Ecole Polytechnique F\'ed\'erale de Lausanne (EPFL),
CH-1015, Lausanne, Switzerland}
\date{\today}

\begin{abstract}

\end{abstract}
\maketitle

\graphicspath{{"d:/Group Vanicek/Desktop/PP_TGA/figures/"}
{./figures/}{C:/Users/Jiri/Dropbox/Papers/Chemistry_papers/2018/PP_TGA/figures/}}

\section{\label{sec:modes}Most displaced modes of the phenyl radical}

\begin{figure}
[H]\centering
\includegraphics[scale=0.07]{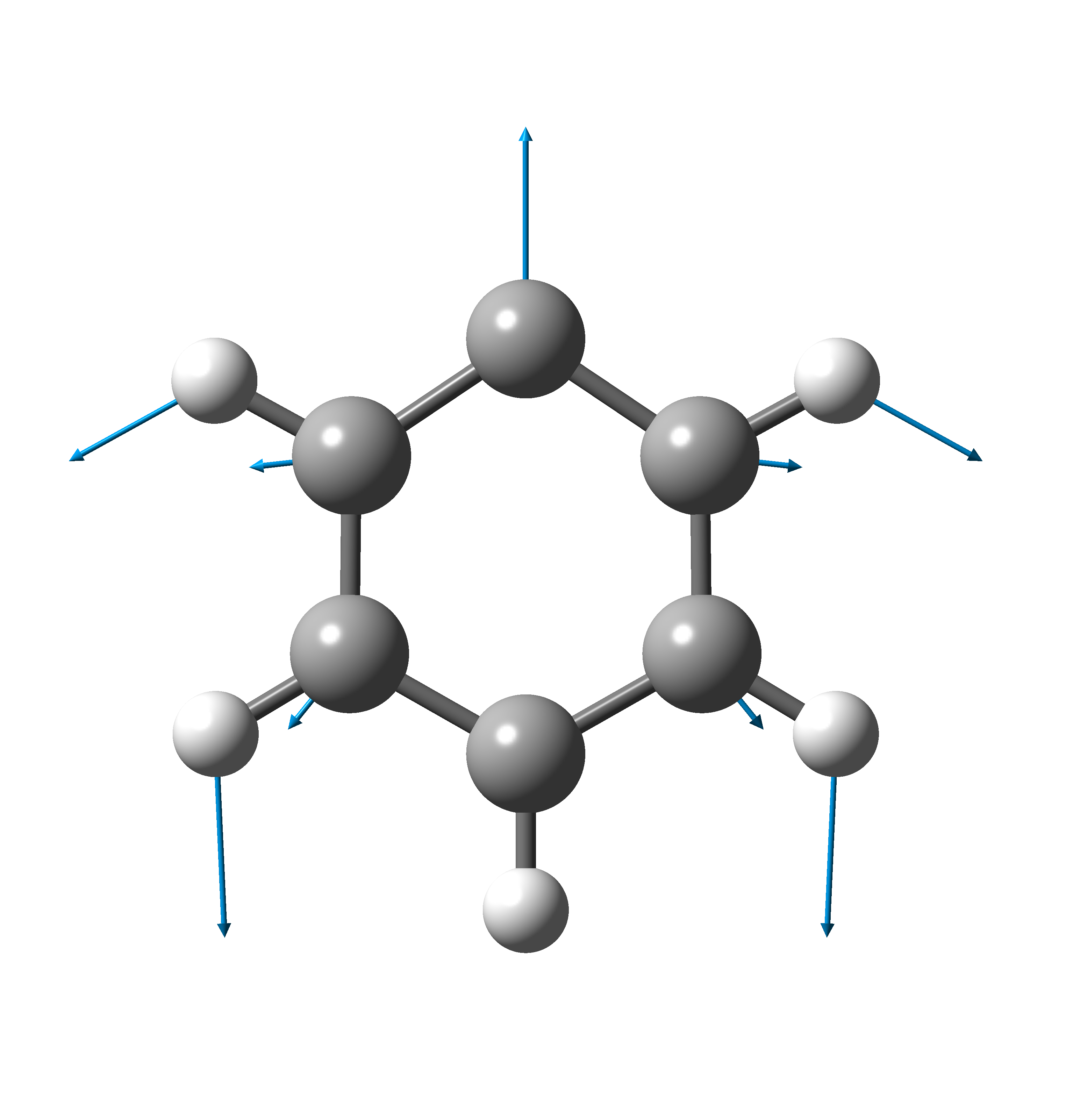} \hspace{1cm}
\includegraphics[scale=0.07]{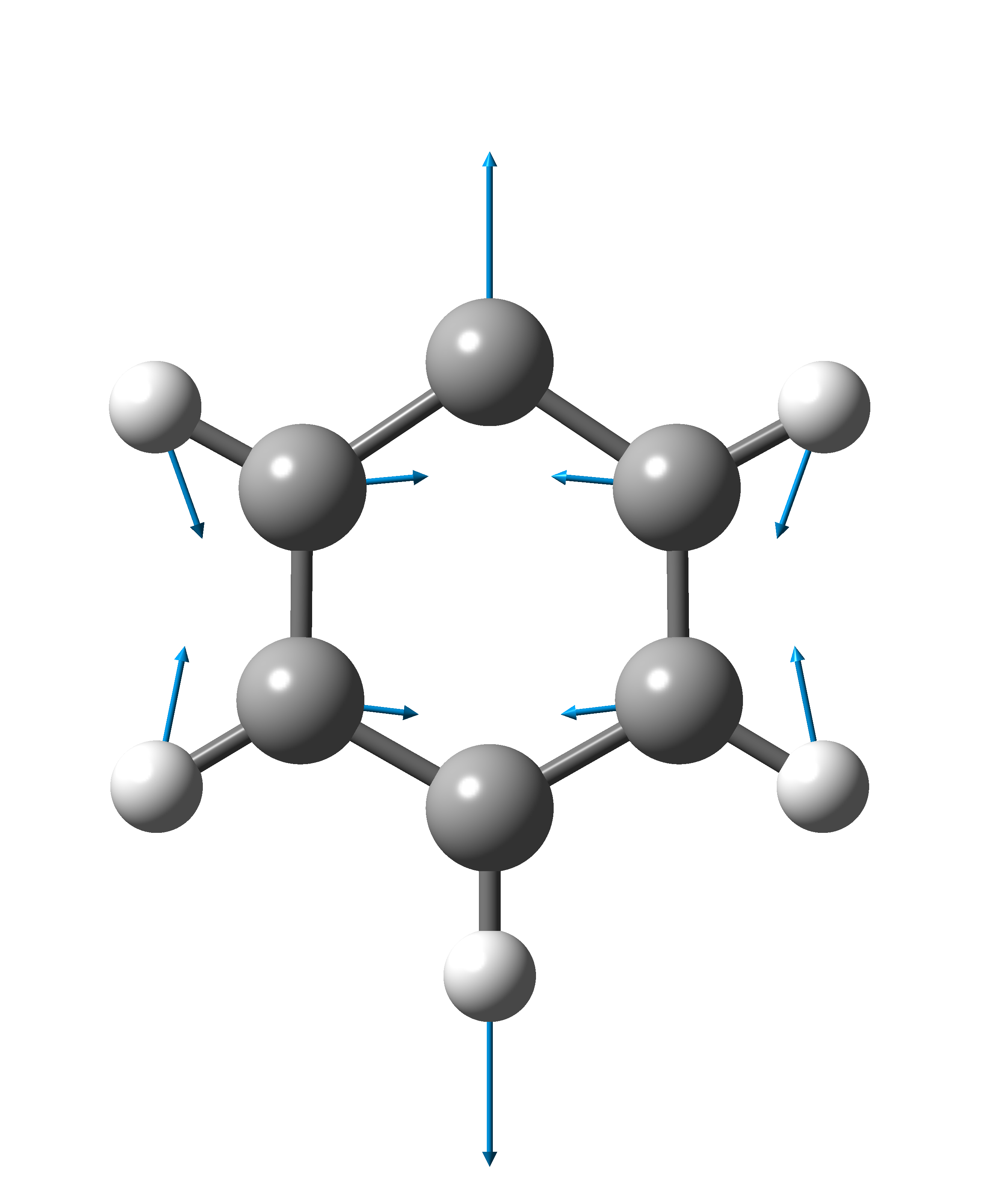} \caption{\label{fig:modes}Most displaced modes of the phenyl radical corresponding to the totally symmetric in-plane vibrations; we show the eigenvectors of the excited-state Hessian evaluated at the optimized geometry of the excited electronic state (\emph{ab initio} method described in the Section~III of the main text). Left: Mode of frequency 924 cm$^{-1}$. Right: Mode of frequency 590 cm$^{-1}$. The full table of the ground- and excited-state frequencies, along with the relative displacements of the corresponding modes can be found in the Supporting Information of Ref.~\onlinecite{Patoz_Vanicek:2018}.}
\end{figure}

\section{\label{sec:model}A one-dimensional \emph{ab initio} model: details of
the potential fit and exact quantum dynamics calculations}

\begin{figure}
[H]\centering
\includegraphics[scale=0.8]{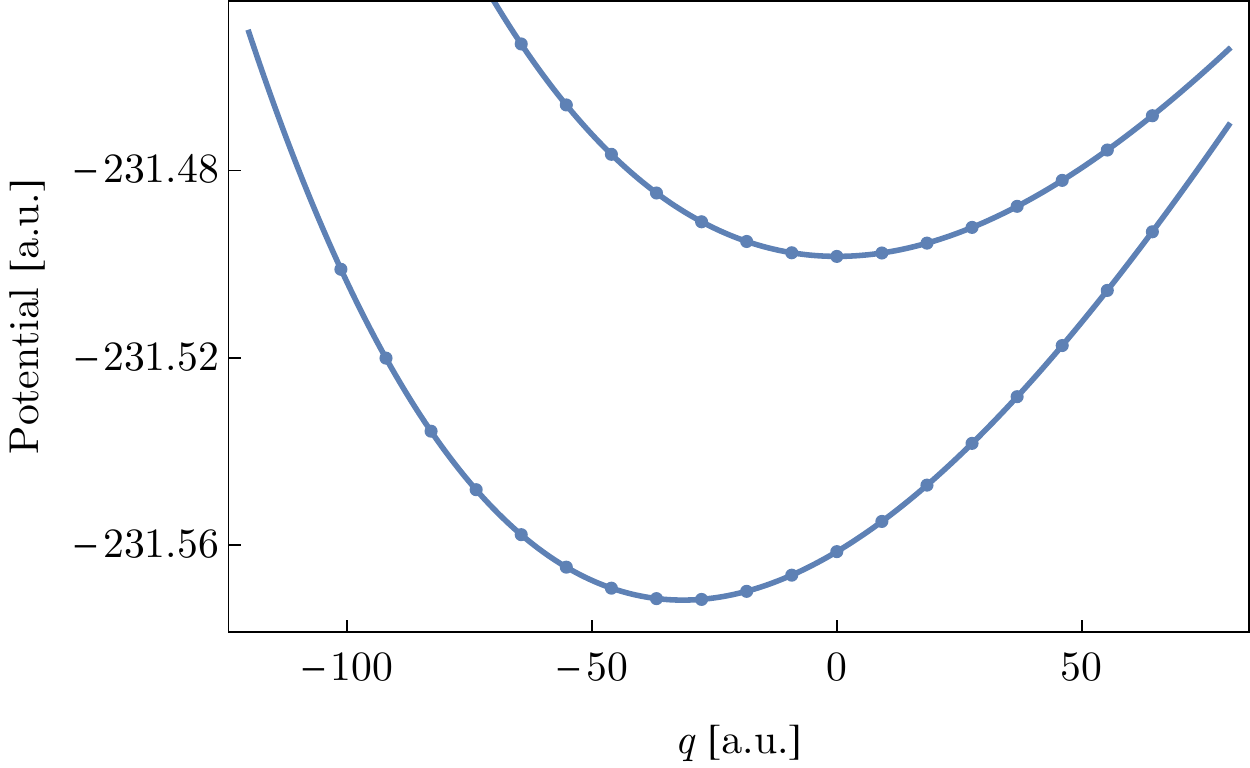} \caption{\label{fig:PotentialPlot}\emph{Ab initio} data (blue points) and the fitted Morse potentials (blue lines) of the ground and excited electronic states of the phenyl radical along the excited-state vibrational mode of frequency 924 cm$^{-1}$; zero position is the minimum of the excited-state potential energy surface (obtained by the \emph{ab initio} geometry optimization).}
\end{figure}

For the model, we chose the excited-state vibrational mode of frequency 924 cm$^{-1}$.
The \emph{ab initio} data was obtained by computing the ground- and
excited-state potentials at the geometries displaced from the excited-state
minimum along the vibrational mode of interest. Then, the data was fitted to a
Morse potential
\begin{equation}
V(q)=V_{0}+d[1-e^{-a(q-q_{0})}]^{2}.\label{eq:Morse}%
\end{equation}
The fitting parameters for the two states are given in
Table~\ref{tab:Morse_Parameters} and the potentials are shown in
Fig.~\ref{fig:PotentialPlot}.

The initial state was obtained as the ground vibrational state of the harmonic
fit to the ground-state Morse potential, i.e., it was a Gaussian with the
initial position corresponding to the ground-state minimum (ground-state
$q_{0}$ in Table~\ref{tab:Morse_Parameters}), zero initial momentum, and
standard deviation $\sigma_{q} = 14.42$ (in atomic units).

\begin{table}
\caption{\label{tab:Morse_Parameters}Parameters (in atomic units) for the Morse potentials [Eq.~(\ref{eq:Morse})] of the ground and excited electronic states.}
\begin{ruledtabular}
\begin{tabular}{cccccccccc}
&& $V_0$ && $q_0$ && $a$ && $d$\\  \hline
Ground state  && -231.57171 && -31.47  && 0.003215 && 1.11757\\
Excited state && -231.49832 && 0.03757 && 0.003173 && 0.88029
\end{tabular}
\end{ruledtabular}

\end{table}

We note that the vibrational modes experience significant coupling and that
the chosen coordinate is not a vibrational mode of the ground-state potential;
a multi-dimensional model would be required to accurately describe the
dynamics of the phenyl radical. Therefore, the initial wavepacket discussed
here, which is obtained from a harmonic fit to the ground-state model
potential, is not physical, i.e., it does not correspond to any of the degrees
of freedom of the full-dimensional initial wavepacket. Nevertheless, the goal
of this calculation was to present the accuracy of the thawed Gaussian
approximation in moderately anharmonic systems by comparing it to the exact
dynamics---we did not aim to recover the spectrum of the phenyl radical. Yet,
our model was based on the \emph{ab initio} data, and so has a realistic
extent of anharmonicity.

Exact quantum dynamics was performed using the second-order split-operator
method; the same time step and total propagation times were used as for the
thawed Gaussian propagation (see Section~III. of the main text). The spatial
grid consisted of $16384$ equidistant points between -200 and 160 atomic units.

\section{\label{sec:absspec}Excited-state wavepacket and absorption spectrum
of the Morse model}

\begin{figure}
[H]\includegraphics[scale=1]{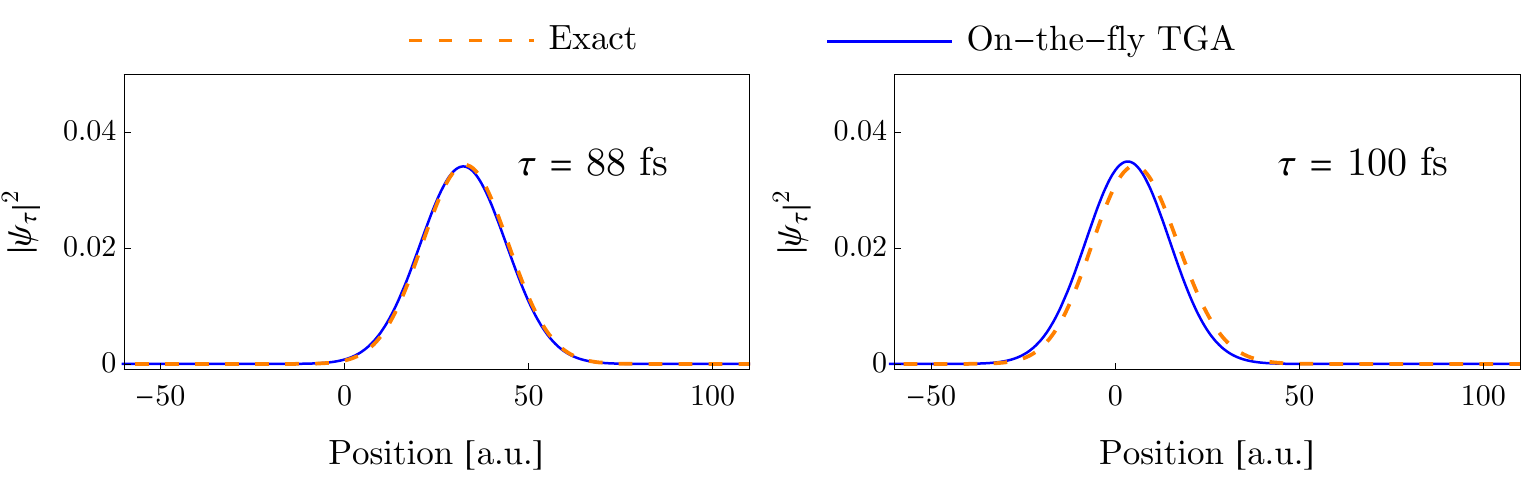}
\caption{\label{fig:WavefunctionExact} Probability density of the excited-state wavepacket at two different probe
delay times $\tau$.}
\end{figure}

\begin{figure}
[H]\includegraphics[scale=1]{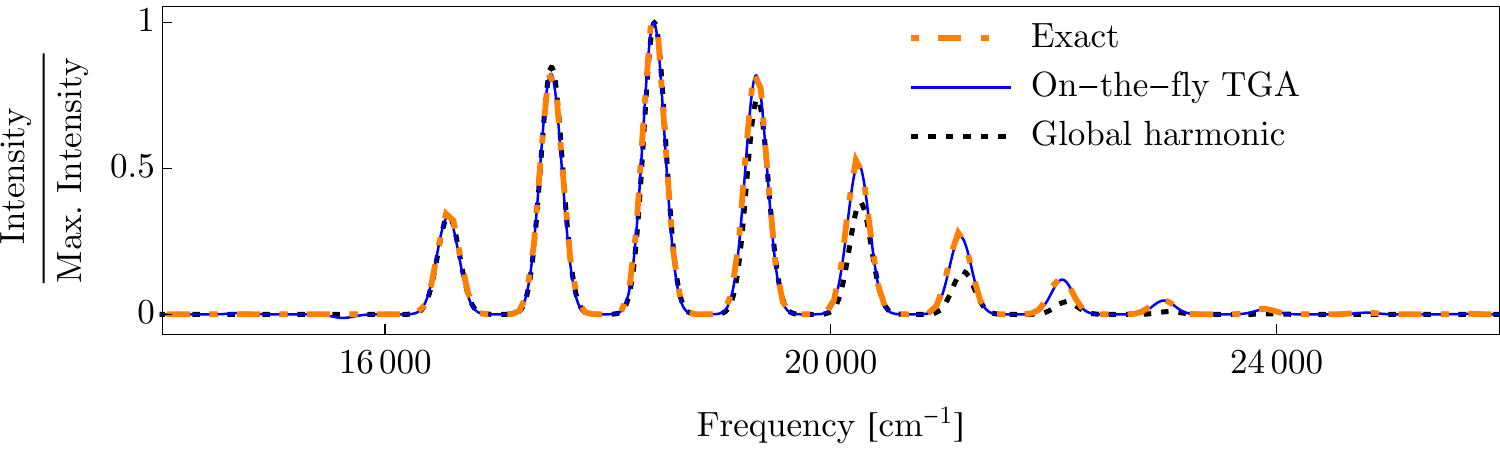}
\caption{\label{fig:SpectrumExact}Linear absorption spectrum of the Morse model described in
Section~\ref{sec:model} of the Supporting Information evaluated using the exact quantum dynamics,
on-the-fly thawed Gaussian approximation (TGA), and global harmonic approach.}
\end{figure}

\section{\label{sec:exagerated_model}Time-resolved spectra, excited-state
wavepacket, and absorption spectrum of the model with increased anharmonicity}

Here, we compare the thawed Gaussian approximation with the exact result for a
more challenging Morse potential, which is constructed from the \emph{ab
initio} model described in the Supporting Information Section~\ref{sec:model}
by increasing the anharmonicity. This is accomplished by modifying the $a$ and
$d$ parameters so that the product $2da^{2}$, equal to the force constant of
the global harmonic approximation at the minimum of the true potential,
remains unchanged. In this example, we multiply the ground- and excited-state
$a$ parameters by 2 and divide the $d$ parameters by 4 (the original
parameters are given in Table~\ref{tab:Morse_Parameters}).

\begin{figure}
[H]\centering\includegraphics[scale=1]{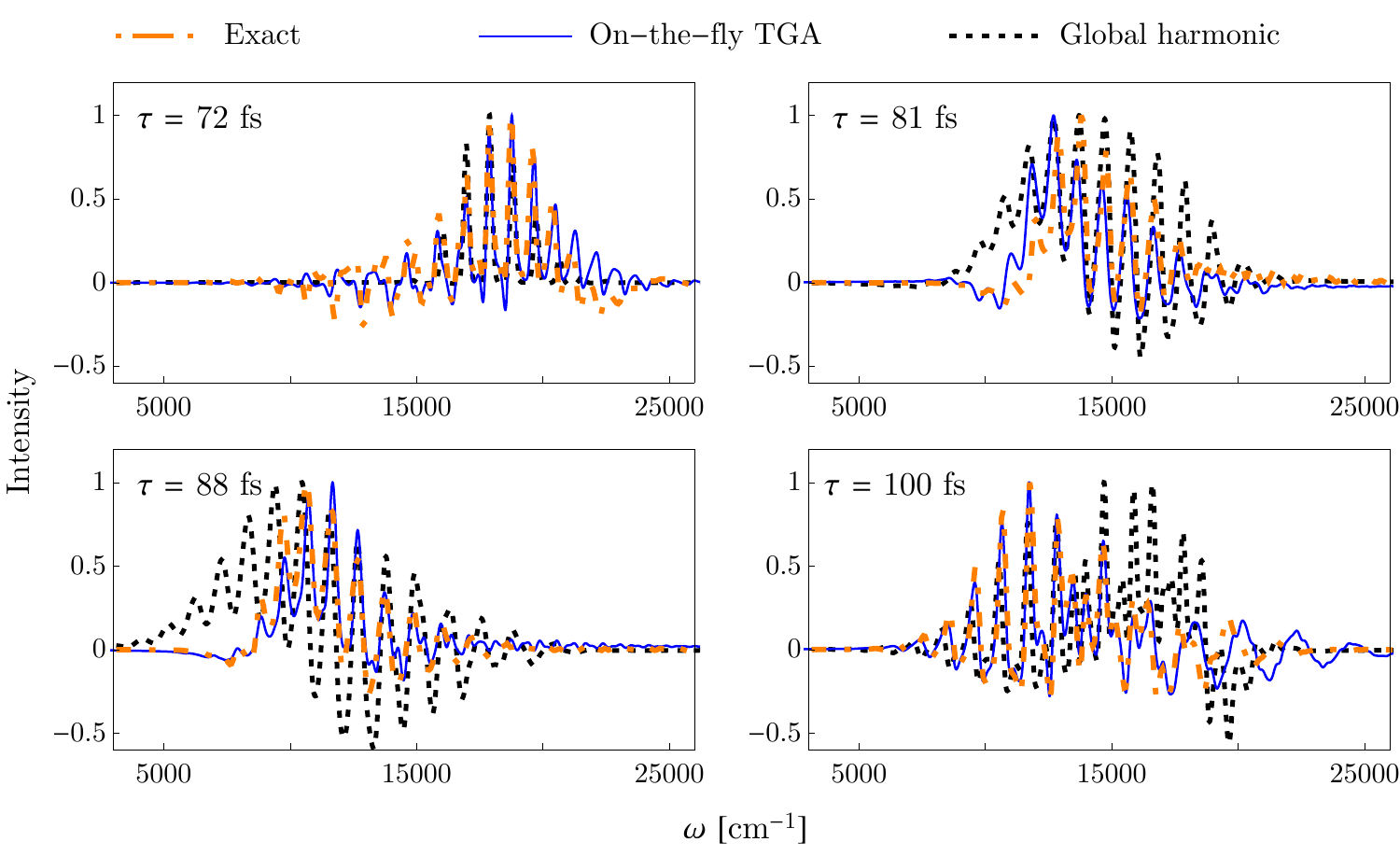}
\caption{\label{fig:ProbesExactExagerated} Time-resolved spectra of the modified Morse potentials with increased
anharmonicity, defined in Section~\ref{sec:exagerated_model} of the Supplementary Information. Analogue of
Fig. 6 of the main text.}
\end{figure}

\begin{figure}
[H]\includegraphics[scale=1]{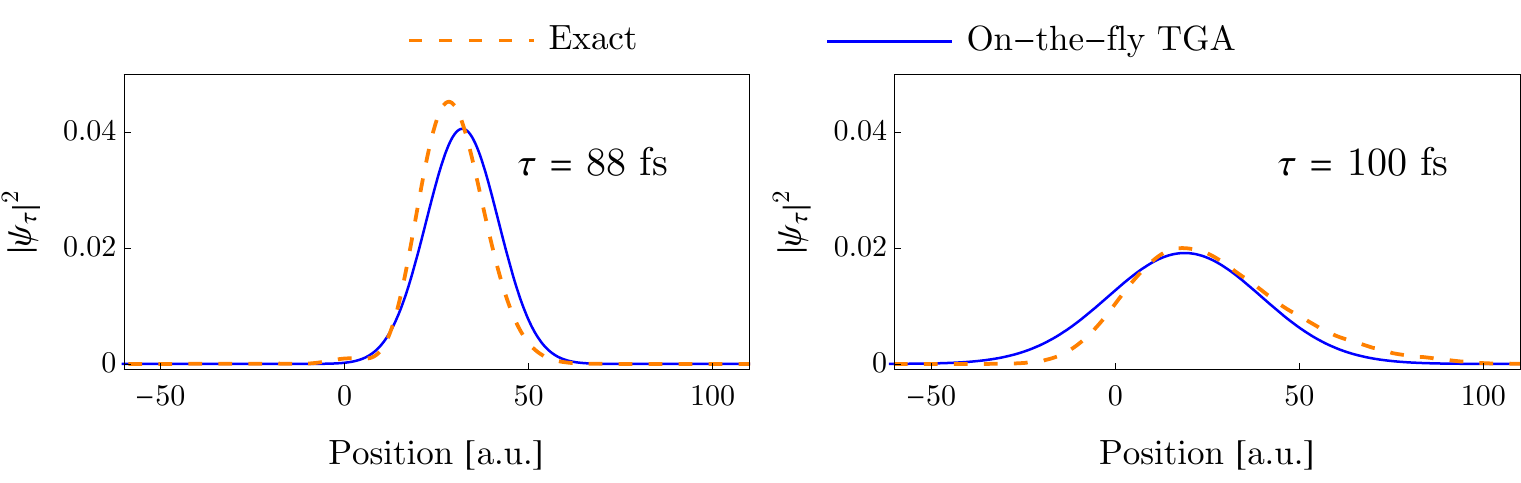}
\caption{\label{fig:WavefunctionExactExagerated} Probability density of the excited-state wavepacket at two
different probe delay times $\tau$. Analogue of Fig.~\ref{fig:WavefunctionExact} of Supporting Information, but
for the modified Morse potentials with increased anharmonicity.}
\end{figure}

\begin{figure}
[H]\includegraphics[scale=1]{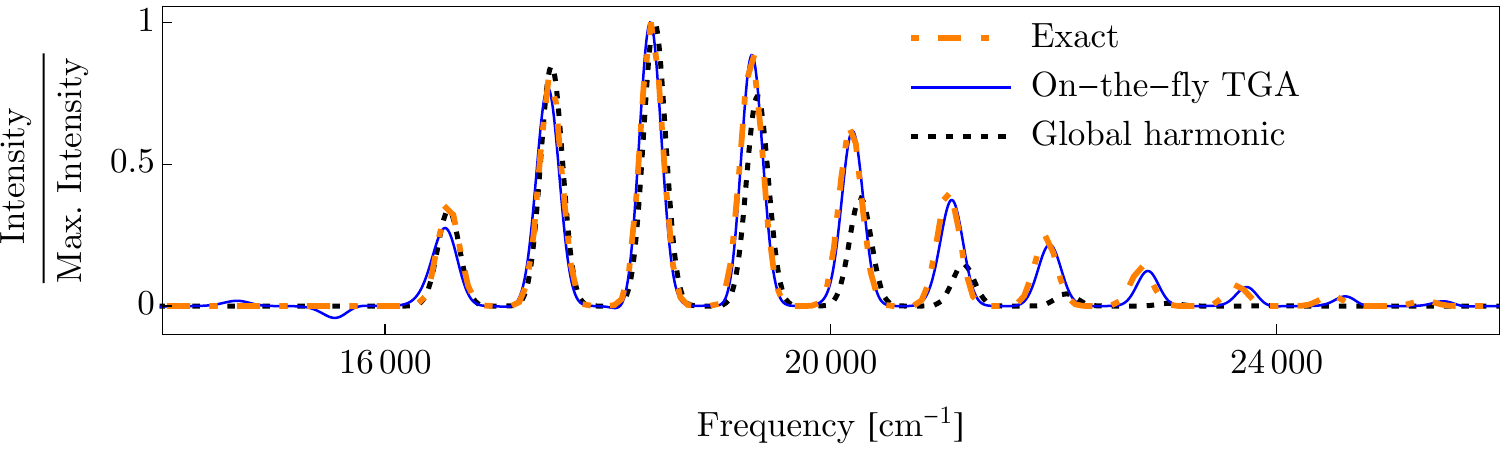}
\caption{\label{fig:SpectrumExactExagerated} Linear absorption spectrum of the modified Morse model with
increased anharmonicity, described in Section~\ref{sec:exagerated_model} of the Supporting Information.
Analogue of Fig.~\ref{fig:SpectrumExact} of Supporting Information.}
\end{figure}

\bibliographystyle{aipnum4-1}
\bibliography{Pump_probe}